\documentclass[aps,prl,twocolumn,amsmath,amssymb,floatfix,superscriptaddress]{revtex4-2}

\usepackage[latin9,utf8]{inputenc}
\usepackage{stmaryrd}
\usepackage{graphicx}
\usepackage{multirow}
\usepackage{mathtools}
\usepackage{xcolor}
\usepackage{textcomp}
\usepackage{gensymb}
\usepackage{calligra}
\usepackage{braket}
\usepackage{float}
\usepackage{bbold}
\usepackage{verbatim}
\usepackage{graphics}

\usepackage{xcolor,mathrsfs,dsfont}
\definecolor{darkblue}{RGB}{0,0,150}
\definecolor{nightblue}{RGB}{0,0,100}
\definecolor{evergreen}{RGB}{0,120,0}
    
\usepackage{graphicx,mathtools,bm}
\graphicspath{{./Figures/}}
\usepackage[
colorlinks,
citecolor=blue,
linkcolor=blue,
urlcolor=nightblue,
breaklinks=true]{hyperref}
\usepackage{breakurl}

\let \oldbm \bm
\renewcommand{\vec}[1]{\oldbm{#1}}

\usepackage[english]{babel}
\usepackage[babel,kerning=true,spacing=true]{microtype}
\usepackage[utf8]{inputenc}

\definecolor{DarkRed}{RGB}{100,0,0}

\def\bm{{\vec m}}

\def\H{\mathcal{H}}

\makeatother

\begin{document}
\title{Quantum metric dependent anomalous velocity in systems subject to complex electric fields}
\author{Bar Alon}
\author{Roni Ilan}
\author{Moshe Goldstein}

\affiliation{Raymond and Beverly Sackler School of Physics and Astronomy, Tel Aviv University, Tel Aviv 6997801, Israel}

\date{\today}

\begin{abstract}
Berry phases have long been known to significantly alter the properties of periodic systems, resulting in anomalous terms in the semiclassical equations of motion describing wave-packet dynamics. In non-Hermitian systems, generalizations of the Berry connection have been proposed and shown to have novel effects on dynamics and transport.
In this work, we consider perturbing fields which are themselves non-Hermitian, in the form of complex external electric fields, which are realizable as gain/loss gradients. We derive the full set of semiclassical equations of motion and show that the anomalous velocity depends not only on the Berry curvature, but on the entirety of the quantum geometric tensor, including the quantum metric. This quantum metric dependent velocity appears regardless of whether the unperturbed Hamiltonian is Hermitian or not.
These analytical results are compared with numerical lattice simulations which reveal these anomalous terms even in one-dimension.
Our work expands the range of phenomena expected to be detectable in experimental setups, which should be realizable in currently available metamaterials and classical wave systems, including mechanical, acoustic, and optical.
\end{abstract}

\maketitle

\section{Introduction} \label{sec:intro}
Semiclassical band theory is the study of the dynamics of wave-packets as they propagate through periodic media. In particular, this theory has been vital in the description of uncorrelated electrons in solids, their transport properties, and their response to external perturbations such as electromagnetic fields.
The semiclassical theory was revolutionized with the introduction of the Berry phase \cite{berryQuantalPhaseFactors1984,ZakPhysRevLett.62.2747}, which is a geometric phase accumulated through the cyclic evolution of the system's parameters. The Berry phase encapsulates geometric and topological properties of the system and is known to have non-trivial effects in a wide variety of fields, including atomic and molecular physics \cite{MeyerPhysRevA.80.062110,MironovaPhysRevA.87.013627}, optics \cite{chiaoBerryPhasesOptics1990,jishaGeometricPhaseOptics2021}, and high-energy physics \cite{StonePhysRevD.91.025004,baggioAspectsBerryPhase2017}. In condensed matter physics, in particular, geometric and topological effects are at the heart of the modern theory of electric polarization \cite{VanderbiltPhysRevB.47.1651,OrtizPhysRevB.49.14202,RestaRevModPhys.66.899}, orbital magnetization \cite{chang_berry_1996,ThonhauserPhysRevLett.95.137205}, and the celebrated quantum Hall effect \cite{ThoulessPhysRevLett.49.405,HaldanePhysRevLett.61.2015}.

Of particular relevance to the semiclassical theory is the Berry curvature, which derives from the Berry phase. This quantity appears directly in the semiclassical equations of motion through the anomalous velocity \cite{chang_berry_1995,sundaram_wave-packet_1999,chang_berry_2008,xiao_berry_2010} and is responsible for topological effects of electronic band theory, with the intrinsic contribution to the anomalous Hall effect \cite{PhysRevLett.88.207208,RevModPhys.82.1539} being chief among them. While the Berry curvature has been studied extensively, it is only one part of a more generic geometric object - the Quantum Geometric Tensor (QGT), also known as the Fubini-Study metric \cite{provost_riemannian_1980,resta_insulating_2011,cayssol_topological_2021}. The rest of the QGT is determined, in turn, by the Quantum Metric (QM). While so far less studied, the QM is nevertheless related to several physical phenomena such as superfluidity in flat bands \cite{LiangPhysRevB.96.064511}, anadiabatic dynamics \cite{Gao_PhysRevLett.112.166601,bleu_effective_2018,Holder_PhysRevResearch.2.033100,leblanc_universal_2021}, current noise spectra \cite{NeupertPhysRevB.87.245103}, and response in driven systems \cite{ozawa_steady-state_2018}. The QM is furthermore vital in determining the real-space spread of Wannier functions \cite{marzari_maximally-localized_1997}.

The applicability of the semicalssical theory is not limited to electronic systems. Growing interest in classical metamaterials has shown that a wide variety of mechanical \cite{VitellichenNonlinearConductionSolitons2014,Roman2015science.aab0239,neder2023bloch}, optical \cite{RaghuPhysRevA.78.033834,wangObservationUnidirectionalBackscatteringimmune2009a,OzawaRevModPhys.91.015006} and acoustic \cite{heAcousticTopologicalInsulator2016,QiPhysRevLett.124.206601} systems exhibit many of the topological and geometric effects of electronic band theory, including Bloch oscillations, Berry phases, and chiral edge modes.
Such classical systems, however, naturally experience dissipation and gain, which in some cases may even be tuned or engineered, leading to their dynamics being described by an effective non-Hermitian Hamiltonian. The full description of such systems, therefore, warrants a novel non-Hermitian topological band theory, which has been a subject of active research in recent years \cite{Bender_2007,ShenPhysRevLett.120.146402,martinezalvarezTopologicalStatesNonHermitian2018,Ghatak_2019,KunstPhysRevB.99.245116,gong_topological_2018,ashida_non-hermitian_2020, Martinezhttps://doi.org/10.1002/qute.202300225}.

The development of such a Non-Hermitian theory, whose results would also be applicable to realizable systems, has proven challenging with many established ideas of the Hermitian theory having to be re-examined. Firstly, the failure of the adiabatic theorem \cite{GNenciu_1992,berry_slow_2011,ibanez_adiabaticity_2014,wang_non-hermitian_2018} for complex spectra leads to instabilities in the single-band picture unless the band in question is maximally amplified, or the spectrum is real, such as in $\mathcal{PT}$-symmetric Hamiltonians \cite{BenderPhysRevLett.80.5243,BENDER20101616}.
Secondly, the possible non-orthogonality of modes leads to many ambiguities in the definitions of the geometric and dynamical variables, such as the Berry connection, band populations and expectation values, leading to multiple and at times conflicting definitions \cite{Brody_2014,ibanez_adiabaticity_2014,ShenPhysRevLett.120.146402}.
Finally, basic ideas of topological band theory such as the bulk-edge correspondence have been known to become much more nuanced or to fail entirely when non-Hermiticity is introduced \cite{bergholtz_exceptional_2021}.

Previous works \cite{graefe_wave-packet_2011,schomerus_non-hermitian-transport_2014,xu_weyl_2017} have had relative success in tackling these challenges by considering a well-localized wavepacket state confined to the maximally amplified band and deriving its dynamics directly through the Schrödinger equation, with some of the novel results already being tested in experimental setups \cite{PhysRevResearch.5.L032026,martello_coexistence_2023}. However, to the best of our knowledge, all previous works of this type only considered non-Hermiticity in the unperturbed Hamiltonian and assume perturbing fields are real.

In this work, we consider a band Hamiltonian perturbed by an external field which is itself non-Hermitian. Specifically, we derive the full set of semi-classical equations of motion for a non-Hermitian Bloch electron subject to a uniform complex electric field. Since such fields correspond to gradients of external gain or dissipation potentials, they are realizable in currently available metamaterial systems.

We derive the existence of new terms in the equations of motion, including a generalized anomalous velocity which couples to the entirety of the QGT. While for real fields \cite{silberstein_berry_2020} this generalized term reduces to the usual anomalous velocity, which depends only on the Berry curvature, this term also depends on the quantum metric for fields with a non-vanishing imaginary part. Although the quantum metric has been known to play a part in wave-packet dynamics in non-Hermitian systems, older works only show its relevance in higher-order effects or near exceptional points \cite{solnyshkov_quantum_2021}. Our work reveals its importance to the dynamics already to first order, and even when the original band Hamiltonian is perfectly Hermitian and no exceptional points exist. Furthermore, our formalism provides a straightforward interpretation of all geometric terms appearing in the equations of motion using the language of non-Hermitian perturbation theory.
We confirm the validity of our results, which produce non-vanishing corrections even for one-dimensional systems, by simulating wave-packet dynamics in both Hermitian and non-Hermitian lattice models.

The remainder of this work is structured as follows: In Sec. \ref{sec:Formalism} we outline the formalism and define the basic quantities of interest. Our main results are presented in Sec. \ref{sec:EOM}, where we begin by deriving general expressions for single-band wavepackets with general perturbations, and then detail the full set of equations of motion for the specific case of a uniform complex electric field. The theoretical results are then tested in Sec. \ref{sec:numerics} using numerical simulations of a 1-dimensional non-Hermitian lattice Dirac model. Finally, we conclude in Sec. \ref{sec:Conclusions}. Detailed derivations of the results, as well as additional discussions relating to the finer points of the simulations are included in the appendices.

\section{Formalism}\label{sec:Formalism}

\subsection{Non-Hermitian systems}

We consider here a physical system whose state is described in general by a wavefunction belonging to some Hilbert space $\ket{\psi}\in\H$ and whose dynamics are governed by a Schrödinger equation 
\begin{equation}
    \frac{d}{dt}\ket{\psi(t)} = -iH\ket{\psi(t)},
\end{equation}
for some Hamiltonian operator $H$ which may not, in general, be Hermitian.

Throughout this work we assume that the Hamiltonian in question has the periodicity of some real-space lattice in the sense that it commutes with a set of lattice-translation operators, and hence Bloch's theorem holds, such that
\begin{equation}
    H\ket{\psi_{nk}}=\varepsilon_{nk}\ket{\psi_{nk}},\quad \psi_{nk}(r)=e^{ikr}u_{nk}(r),
\end{equation}
where the $u_{nk}(r)$ have the periodicity of the lattice. The careful reader will note that since $H$ is not assumed to be Hermitian it may not necessarily be diagonalizable, and hence a complete set of $\ket{\psi_{nk}}$ may not exist. However, it is still always possible to choose a \emph{maximal} set of eigenstates of $H$ such that each state has a well defined Bloch momentum $k$. Furthermore, these momenta are good quantum numbers, i.e. preserved by the dynamics.

We turn now to the discussion of the various unusual phenomena which arise due to the non-Hermitian nature of $H$. Firstly, the spectrum of $H$ may no longer be real, giving rise to complex energies. Physically, the imaginary part of the energy expresses the gain (loss) of its associated mode, meaning that the norm of the mode is not necessarily conserved.
More importantly, since different modes experience different gains (losses), the different components of a superposition of eigenmodes will be amplified (decay) at different rates. As a result, the amplitude of the maximally amplified mode in any given superposition grows exponentially in time relative to the other modes, causing the amplitude of the state to become increasingly concentrated in the maximally amplified mode until the weight of all other modes in the superposition becomes negligible. This effect has profound consequences on the applicability of the adiabatic theorem as even very weak band transitions induced by slow perturbations may be exponentially amplified over time \cite{wang_non-hermitian_2018,silberstein_berry_2020}.

Secondly, the eigenmodes of the Hamiltonian may not be orthogonal to each other, even if they do still form a complete basis. This occurs whenever the Hamiltonian fails to be \emph{normal} i.e. it does not commute with its Hermitian conjugate: $[H,H^\dagger]\neq0$. In such cases, it is convenient to term the eigenbasis of the Hamiltonian as the \emph{right} eigenbasis $\ket{\psi^R_{nk}}$ and define the \emph{left} eigenbasis $\ket{\psi^L_{nk}}$ to be its dual, or reciprocal, basis
\begin{equation}
    \label{eq:def_leftright}
    \braket{\psi^L_{n^\prime k^\prime} | \psi^R_{nk}} = \delta_{nn^\prime}\delta(k-k^\prime).
\end{equation}
As has already been recognized \cite{schomerus_nonreciprocal_2020}, and will be discussed further below, the left eigenbasis is crucial to describing the response of the system to external perturbations, giving the non-orthogonality of modes far-reaching consequences on the system's dynamics and behaviour.

For a system with non-orthogonal modes, it is also useful to define the Gram matrix of the eigenbasis:
\begin{equation}
    \label{eq:grammian}
    I_{n^\prime k^\prime;nk} = \braket{\psi^R_{n^\prime k^\prime} | \psi^R_{nk}} = I_{n^\prime n}(k)\delta(k-k^\prime).
\end{equation}
Note that, due to the symmetries assumed for our Hamiltonian, $I$ is diagonal with respect to $k$ since it is still a good quantum number defined by a Hermitian operator, meaning that while different bands may not be orthogonal to each other for a given value of $k$, states with different momenta must remain orthogonal. In the familiar case of Hermitian (and hence normal) Hamiltonians, the eigenbasis may always be chosen to be orthonormal and hence the left and right eigenbases become identical and $I$ is nothing more than the identity matrix.

Lastly, the Hamiltonian may fail to be diagonalizable at all. For a Bloch Hamiltonian $H(k)$ this occurs at degeneracy points where not only eigenvalues, but also eigenvectors coalesce. Such points are known as `exceptional points' and have been shown to relate to the topological properties of the system and exhibit various extreme effects \cite{xu_weyl_2017,AliMiriExceptional,solnyshkov_quantum_2021,bergholtz_exceptional_2021}. In this work, however, we will assume that we are well away from such points such that diagonalizability always holds .

We define here the expectation value of a general state $\ket{\psi}$ as
\begin{equation}
    \label{eq:def_expectation}
    \braket{\hat{O}}_\psi=\frac{\bra{\psi}\hat{O}\ket{\psi}}{\braket{\psi|\psi}}.
\end{equation}
While alternative definitions for the expectation value exist in the non-Hermitian setting, this definition is unique in encapsulating the idea that physical observables are functions of state in the sense that the expectation of a state $\ket{\psi(t)}$ at a particular moment in time is completely independent of the dynamics of the system before and after that moment \cite{silberstein_berry_2020}.

We define now the basic geometric quantities of interest. Firstly, we present two generalizations of the Berry connection \cite{ShenPhysRevLett.120.146402}:
\begin{equation}
    \label{eq:Berry_connections}
    A^{RR}_\lambda = i\frac{\braket{u^R_{nk}|\partial_\lambda u^R_{nk}}}{I_{nn}(k)},\quad A^{LR}_\lambda = i\braket{u^L_{nk}|\partial_\lambda u^R_{nk}},
\end{equation}
with $\lambda$ some arbitrary parameter (which may or may not be a crystal momentum).
The first generalization, $A^{RR}$, is nothing more than the normalized Berry connection of the functions which make up the $n$th band. Like its Hermitian counterpart, this connection is real if the gauge is chosen such that the cell-periodic functions all have the same norm.
The second generalization, $A^{LR}$, involves left states and hence is not a property of the band itself. This quantity, like others involving left states, relates to the response of the system to perturbations \cite{schomerus_nonreciprocal_2020} and has been shown to play a role in the dynamics of perturbed systems through the gauge-invariant difference $A^{RR}-A^{LR}$ \cite{xu_weyl_2017,silberstein_berry_2020,PhysRevResearch.5.L032026}. The contour integral of $A^{LR}$ yields the non-Hermitian version of the Berry phase \cite{GDattoli_1990,longhi_complex_2023}.

It is also useful to define the QGT of the band
\begin{equation}
    \label{eq:def_QuantumGeometricTensor}
    T_{\lambda\mu} = \frac{\braket{\partial_\lambda u^R_{nk} | \partial_\mu u^R_{nk}}}{I_{nn}(k)} - \frac{\braket{\partial_\lambda u^R_{nk}| u^R_{nk}} \braket{u^R_{nk}|\partial_\mu u^R_{nk}}}
    {I^2_{nn}(k)}. 
\end{equation}
This is a straight-forward generalization of the quantum geometric tensor which appears in the Hermitian setting \cite{provost_riemannian_1980,resta_insulating_2011,cayssol_topological_2021}. We note here that this tensor is gauge-invariant and involves only eigenstates of a single band and their derivatives.

Similarly to the Hermitian case, the QGT is represented by a Hermitian matrix and may be decomposed into its real (symmetric) and imaginary (anti-symmetric) components
\begin{equation}
    \label{eq:QGT_split}
    T_{\lambda\mu} = g_{\lambda\mu} - \frac{i}{2}\Omega_{\lambda\mu},
\end{equation}
with the quantum metric $g_{\lambda\mu}$ being the real part and the Berry curvature $\Omega_{\lambda\mu}$ being proportional to the imaginary part. Both parts are thus real and gauge-invariant by definition. Note that we have elected to omit the superscript $RR$ from our notation of the Berry curvature and the QGT as, while other versions of these quantities which involve left eigenstates have been defined \cite{ZhangPhysRevA.99.042104, ZhuPhysRevB.104.205103}, they do not play a part in the dynamics, and thus we do not make use of them in this work. In complete analogy to the Hermitian setting, the Berry curvature also arises as the curl of the appropriate Berry connection
\begin{equation}
    \label{eq:def_Bcurvature}
    \Omega_{\lambda\mu} = \partial_\lambda A^{RR}_\mu - \partial_\mu A^{RR}_\lambda.
\end{equation}
We emphasize that this implies that the non-Hermitian Berry phase of an adiabatic evolution is \emph{not} given by an integral of this Berry curvature over a surface suspending the adiabatic loop in parameter space, as they emerge from
different Berry connections.

As is well known from the Hermitian setting, the Berry curvature quantifies the geometric response of the system to perturbations and is responsible for the anomalous velocity that arises in external electric fields. As we show here, however, once the system is subject to a non-Hermitian perturbation, e.g. a complex electric field, the entirety of the quantum geometric tensor may play a role in the dynamics.

\subsection{Non-Hermitian perturbation theory}
Since we are interested in the response of the non-Hermitian system to external fields, it is useful to consider the problem through a non-Hermitian generalization of perturbation theory. Such a generalization may be readily obtained by considering some non-Hermitian parameterized Hamiltonian $H(\lambda)$ through the Hellmann-Feynman approach. The first order results of this theory are \cite{sternheim_non-hermitian_1972}
\begin{equation}
\label{eq:perturbation_energy}
    \partial_\lambda \varepsilon_n = \bra{n^L}\partial_\lambda H\ket{n^R},
\end{equation}
\begin{equation}
\label{eq:perturbation_state}
    \ket{\partial_\lambda n^R} = \sum_{m\neq n}\frac{\bra{m^L}\partial_\lambda H\ket{n^R}}{\varepsilon_n-\varepsilon_m}\ket{m^R} - iA^{LR}_\lambda\ket{n^R}.
\end{equation}
These equations, while simple, are difficult to apply directly to Bloch states perturbed by external fields simply because the expectation value of the position operator is ill-defined when applied to the extended Bloch states.

In the Hermitian setting, this difficulty is overcome by replacing the Bloch state $\ket{\psi_{nk}}$ with a localized wavepacket that is sharply peaked in momentum space and taking the limit as the packet tends towards the Bloch state of interest. The non-Hermitian case is more delicate, however, as \eqref{eq:perturbation_energy}, \eqref{eq:perturbation_state} involve `bi-orthogonal' matrix elements which connect left eigenstates with right eigenstates. This again produces some difficulty as while the elements of the left eigenbasis are well defined there is no meaningful way to define the left (or reciprocal) counterpart of a general state, and by extension of the localized wavepacket.

We solve this difficulty by defining the band projection operator
\begin{equation}
\label{eq:def_Pn}
    P_n= \int_k \ket{\psi^R_{nk}}\bra{\psi^L_{nk}}.
\end{equation}
This is indeed a projection operator which annihilates all but the $n$th band of $H$ and it is Hermitian exactly when the left and right eigenstates of the band coincide. That is, when the $n$th band is orthogonal to all others.

The importance of the projector comes about in re-defining the bi-orthogonal matrix element in the $n$th band as an expectation value
\begin{equation}
\label{eq:perturb_withproj}
    \bra{\psi^L_{nk}}\partial_\lambda H\ket{\psi^R_{nk}} = \frac{\bra{\psi^R_{nk}}P_n\partial_\lambda H\ket{\psi^R_{nk}}}{\braket{\psi^R_{nk}|\psi^R_{nk}}}.
\end{equation}
Being an expectation value of well defined operators, this re-formulated definition may be applied to localized wavepacket states restricted to the $n$th band and approximating $\ket{\psi^R_{nk}}$. As we show in section \ref{sec:EOM}, the energy corrections predicted by this definition are exactly those which appear in the semiclassical equations of motion.

\section{Equations of motion} \label{sec:EOM}
\subsection{General formulation of wave-packet dynamics}
We begin by considering a wave-packet state constructed from eigenfunctions of the bare Hamiltonian belonging to a single band $n$
\begin{equation}
    \ket{W}=\int_k w_{nk}\ket{\psi^R_{nk}}.
\end{equation}

We require that this state be well-localized in momentum space in the sense that the envelope function, defined here as
\begin{equation}
    \label{eq:def_envelope}
    g(k)=\vert w_{nk}\vert^2I_{nn}(k),
\end{equation}
is sharply peaked around some central momentum $k_c$
\begin{equation}
\label{eq:peakCondition}
    g(k) \sim N\delta(k-k_c),
\end{equation}
where $N$ is the squared norm of the packet $\braket{W|W}$. We note here that this definition of the envelope function, unlike ones which exclude the Gram matrix $I$, is invariant under renormalizations of the basis functions $\ket{\psi_{nk}^R}$.

If the wave-packet is confined to the \emph{maximally amplified} band, then it adheres to the non-Hermitian adiabatic theorem \cite{GNenciu_1992,silberstein_berry_2020} and we may write, even in the presence of a small perturbation, the Schrödinger equation in the single-band approximation
\begin{equation}
    \dot{w}_{nk} = -i\int_{k^\prime}w_{nk^\prime} \bra{\psi^L_{nk}}H\ket{\psi^R_{nk^\prime}},
\end{equation}
which is equivalent to the usual Schrödinger equation with a band-projected Hamiltonian $P_nH$, where $H=H_0+V$ is some weakly perturbed band Hamiltonian $H_0$ whose solution is known.

By applying this Schrödinger equation to the normalized time derivative of an observable matrix element, we derive a single-band version of the non-Hermitian Ehrenfest theorem
\begin{equation}
    \frac{\partial_t\bra{W}\hat{O}\ket{W} }{\braket{W|W}}=
    \braket{\partial_t \hat{O}}_W - i\braket{\hat{O}P_nH-H^\dagger P_n^\dagger\hat{O}}_W,
\end{equation}
where $\braket{\cdot}_W$ signifies the expectation value with respect to the state $\ket{W}$.
In the case where $\hat{O}$ is time-independent and Hermitian, this simplifies to
\begin{equation}
    \label{eq:SimpleEhrenfest}
    \frac{\partial_t\bra{W}\hat{O}\ket{W} }{\braket{W|W}}=
    2\text{Im}\left( \braket{\hat{O}P_nH}_W \right).
\end{equation}
We may now obtain a general expression for the relative norm-growth simply by setting $\hat{O}$ to be the identity operator and recalling that, by definition, $P_n$ commutes with $H_0$, and obtain
\begin{equation}
    \label{eq:Ndot_generic}
    \frac{\dot{N}}{N}=2\text{Im}\left( \braket{H_0}_W +\braket{P_nV}_W\right),
\end{equation}
which is nothing more than twice the imaginary part of the energy, perturbed to first order, as predicted by \eqref{eq:perturb_withproj}. We stress here that, although we focus on electric fields in this work, the above conclusion is entirely general, and holds for \emph{any} general perturbation $V$, be it an electric field or otherwise, so long as the single-band approximation holds.
From equations \eqref{eq:SimpleEhrenfest} and \eqref{eq:Ndot_generic} we may calculate the equation of motion for any Hermitian operator
\begin{equation}
    \label{eq:SB_Ehrenfest}
    \frac{d}{dt}\braket{\hat{O}}_W =
    2\text{Im}\left( \braket{\hat{O}P_nH}_W \right)
    -\frac{\dot{N}}{N}\braket{\hat{O}}_W.
\end{equation}
This Ehrenfest-like result is again entirely general, and may be applied to any operator, and any perturbed band Hamiltonian $H=H_0+V$, so long as the single-band approximation holds. In this work, this result is the basis from which we derive the equations of motion in a constant complex electric field.

The results derived for Hermitian operators throughout this section are easily generalized to non-Hermitian ones by noting that any operator may be decomposed into a Hermitian and an anti-Hermitian part
\begin{subequations}
\begin{align}
    \hat{O} &= \hat{O}_r+i\hat{O}_i,\\
    \hat{O}_r&=\frac{\hat{O}+\hat{O}^\dagger}{2},\\
    \hat{O}_i&= \frac{\hat{O}-\hat{O}^\dagger}{2i}.
\end{align}
\end{subequations}
Since $\hat{O}_r$, $\hat{O}_i$ are Hermitian, one may apply the above results to each of them separately and then make use of linearity to arrive at the result for the general operator $\hat{O}$.

\subsection{EOM for complex electric fields}
We consider now a (not necessarily Hermitian) band Hamiltonian perturbed by a constant electric field which may in general be complex
\begin{equation}
    \label{eq:ElecPotential}
    V=eE\cdot(\hat{r}-r_c)-e\Phi(r_c),
\end{equation}
where $E$ is the electric field and $\Phi(r_c)$ is the electric potential at the wave-packet center-of-mass. Note that, since a uniform imaginary potential expresses uniform gain (loss) in the system, which is a physical effect, a complex constant electric potential may \emph{not}, in general, be gauged out, in contrast to a purely real field. The importance of this fact is made explicit by the inclusion of $\Phi(r_c)$ in the above expression for the perturbative potential.  

The relative norm growth is then simply given by equation \eqref{eq:Ndot_generic}
\begin{equation}
    \label{eq:Wexp_N}
    \frac{\dot{N}}{N}=2\text{Im}\left(\braket{H_0}_W+eE\cdot\braket{P_n(\hat{r}-r_c)}_W-e\Phi(r_c)\right).
\end{equation}
The above expectation values may be calculated explicitly (see Appendix \ref{AppEOM}) to arrive at
\begin{equation}
\begin{split}
    \label{eq:EOM_N}
    \frac{\dot{N}}{N}=
    &2\text{Im}\left(\varepsilon_{nk}+eE\cdot(A^{LR}-A^{RR})\right)\\
    &-2\text{Im}\left(e\Phi(r_c)\right),
\end{split}
\end{equation}
where the first term encapsulates the perturbative response of the system to the electric field due to the non-orthogonality of its modes, and the second term is simply the gain generated by the external field at the wave-packet center.

The equation of motion for the momentum may now be calculated using \eqref{eq:SB_Ehrenfest}
\begin{equation}
    \label{eq:EOM_k}
    \dot{k} = -\text{Re}\left( eE\right).
\end{equation}
From this result we conclude that while a real field generates a force, a purely imaginary one does not, as it only involves position-dependent gain and does not generate any spectral flow.

We may similarly use \eqref{eq:SB_Ehrenfest} to derive an expression for the velocity
\begin{equation}
    \label{eq:EOM_r}
    \begin{split}
            \dot{r}_i=&
            \frac{\partial }{\partial k_i}\text{Re}\left( \varepsilon_{nk}+eE\cdot(A^{LR}-A^{RR}) \right)\\
            &+ 2\text{Im}(eE_j)\Sigma_{ij}\\
            &- 2\text{Im}\left( eE_jT_{ij} \right),
    \end{split}
\end{equation}
where $T_{ij}$ is the quantum geometric tensor defined in \eqref{eq:def_QuantumGeometricTensor} for parameters $k_i$, $k_j$ and $\Sigma$ is the wave-packet covariance matrix defined as
\begin{equation}
    \label{eq:def_Cov}
    \Sigma_{ij}=\braket{\hat{r}_i\hat{r}_j}_W-\braket{\hat{r}_i}_W\braket{\hat{r}_j}_W.
\end{equation}
The first term of \eqref{eq:EOM_r} is simply the momentum space derivative of the real part of the energy, including the perturbation, and encapsulates the usual group velocity.
The second term arises due to the presence of a gain gradient across the finite width of the packet, which gradually enhances the portions of the packet where the gain is strongest, thereby shifting its center of mass. We name this term the `gain velocity'.
The third term, which we name the `generalized anomalous velocity', reveals a somewhat surprising coupling between the electric field and both the symmetric and anti-symmetric parts of the quantum geometric tensor. Indeed, for purely real fields one recovers the familiar anomalous velocity, which involves only the Berry curvature. This is in agreement with known results for real fields in both Hermitian and non-Hermitian systems \cite{silberstein_berry_2020}.

For complex fields, the anomalous velocity may also involve the quantum metric, which is a unique feature of the complex field. Specifically, for a purely imaginary field $E=i\mathcal{E}$, \eqref{eq:EOM_r} reduces to
\begin{equation}
    \label{eq:EOM_r_imag}
    \begin{split}
            \dot{r}_i=&
            \frac{\partial }{\partial k_i}\left( \text{Re}\varepsilon_{nk}-e\mathcal{E}\cdot\text{Im}(A^{LR}-A^{RR}) \right)\\
            &+ 2e\mathcal{E}_j\Sigma_{ij}\\
            &- 2e\mathcal{E}_jg_{ij}.
    \end{split}
\end{equation}
We emphasize that this novel term is completely independent of the non-Hermitian nature of the unperturbed band Hamiltonian, or lack thereof, and arises purely due to the complexity of the field itself. Furthermore, we notice that since, unlike the Berry curvature, the quantum metric is a symmetric tensor, the generalized anomalous velocity becomes relevant even in one-dimensional systems \footnote{While non-Hermitian systems have already been shown to have corrections to the velocity even in 1-dimension for real electric fields \cite{silberstein_berry_2020}, we now understand them to arise due to a perturbative correction to the band energies, which is separate from the generalized anomalous velocity defined here.}. 

Finally, we note that the generalized anomalous velocity and the gain velocity are not entirely unrelated. It has long been known \cite{marzari_maximally-localized_1997} that the quantum metric is closely related to the localization properties of band-theory wavefunctions. Specifically, the Brillouin zone average of the trace of the metric gives a lower bound on the localization of Wannier functions.
In our setting, it turns out that the difference
\begin{equation} \label{eq:def_CovReduced}
    \overline{\Sigma}_{ij} = \Sigma_{ij}-g_{ij}
\end{equation}
is always positive definite, meaning the quantum metric at the wavepacket center $k_c$ acts as a lower bound for the real-space spread of the packet. While $\overline{\Sigma}$ cannot be evaluated for a general wavepacket, it can be given a closed form that depends only on the shape of the envelope function $g(k)$. These properties are derived in Appendix \ref{App:Cov}.

\section{Numerical Simulations} \label{sec:numerics}

\begin{figure*}[t]
    \centering
    \includegraphics[width=\textwidth]{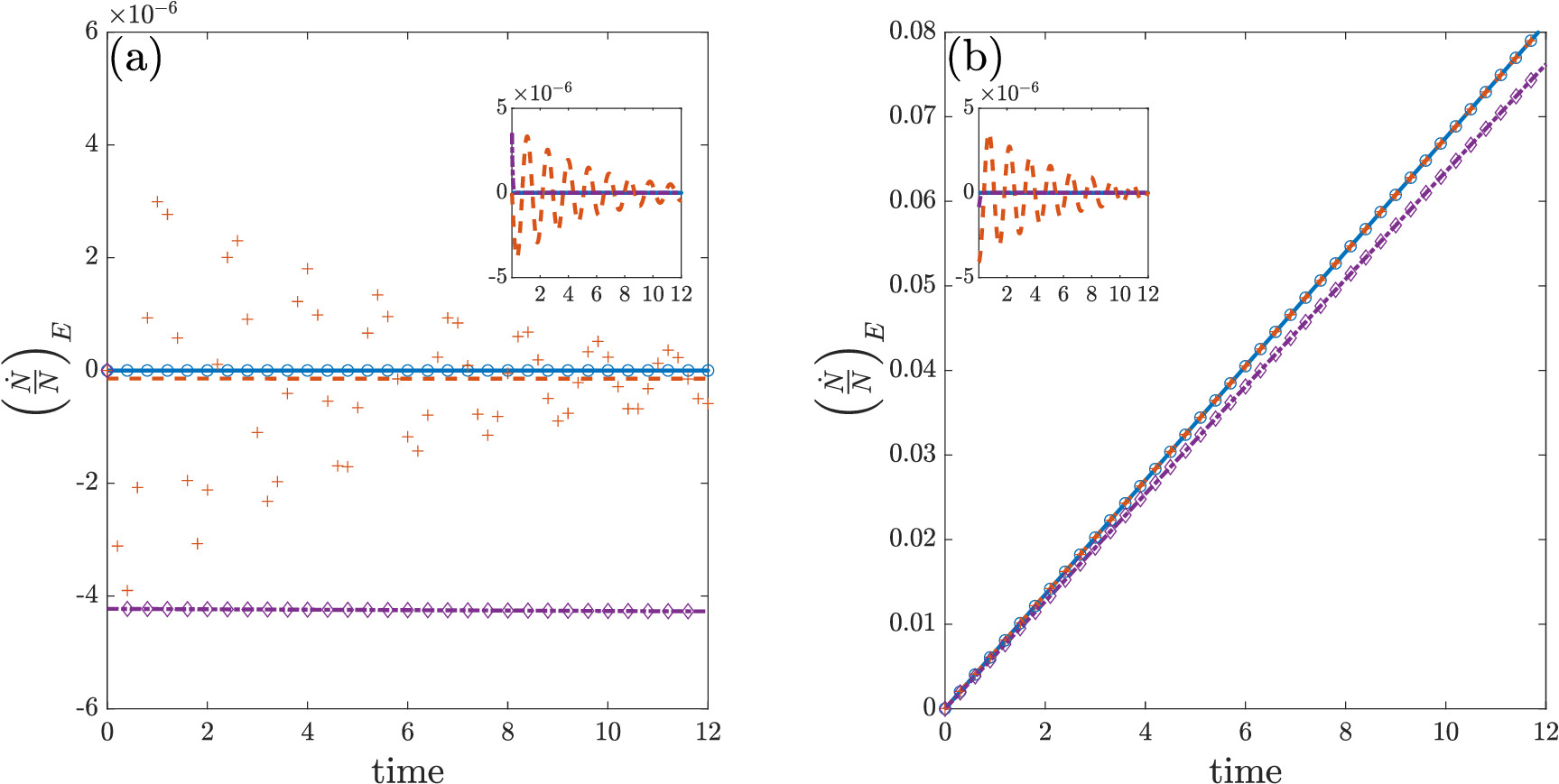}
    \hfill
    \caption{Time evolution of the field-induced norm growth \eqref{subeq:N_E}, for the model parameters summarized in Table \ref{table:modelParams}. Qualitative difference between real and imaginary field is shown. (a) Parameter choices (I)-(III). Simulated results shown in blue circles, orange plus-signs, and purple diamonds, respectively. Theoretical predictions given by \eqref{eq:EOM_N} shown in solid blue line, dashed orange line, and dash-dotted purple line, respectively. (b) Same for parameter choices (IV)-(VI). Time is shown in units where $\hbar=1$. Simulations were conducted with $L=1000$ sites, packet-width $\sigma=0.0126$, and initial momentum $k_c=1$. The timestep used for each simulation was $0.01/\vert m\vert$. Shown in the insets is the difference between the simulation and the theoretical prediction.}
    \label{fig:N}
\end{figure*}

\begin{figure*}[t]
    \centering
    \includegraphics[width=\textwidth]{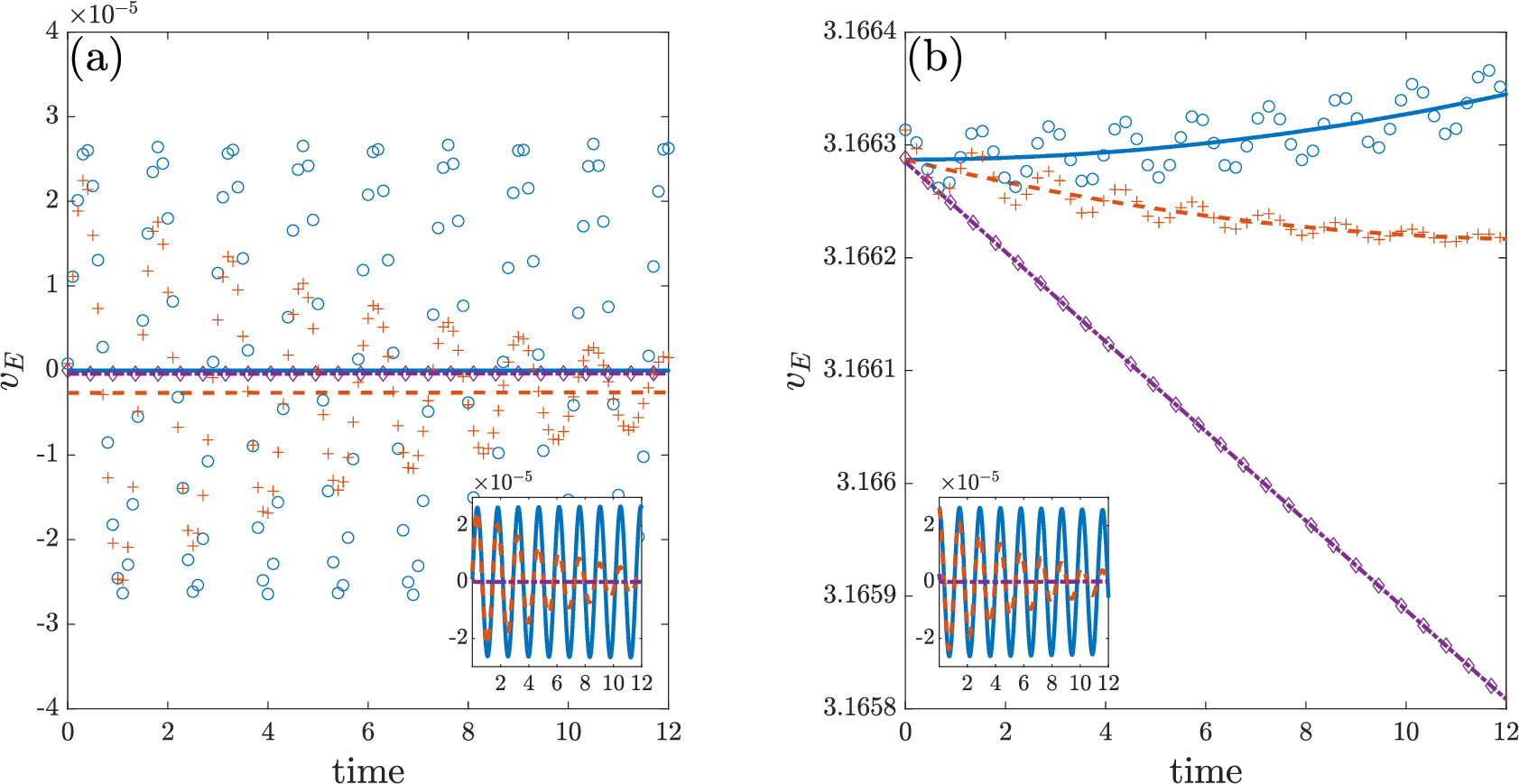}
    \hfill
    \caption{Time evolution of the field-induced velocity \eqref{subeq:v_E}, for the model parameters summarized in Table \ref{table:modelParams}. Qualitative difference between real and imaginary field is shown. (a) Parameter choices (I)-(III). Simulated results shown in blue circles, orange plus-signs, and purple diamonds, respectively. Theoretical predictions given by \eqref{eq:EOM_r} shown in solid blue line, dashed orange line, and dash-dotted purple line, respectively. (b) Same for parameter choices (IV)-(VI). Time is shown in units where $\hbar=1$. Simulations were conducted with $L=1000$ sites, packet-width $\sigma=0.0126$, and initial momentum $k_c=1$. The timestep used for each simulation was $0.01/\vert m\vert$. Shown in the insets is the difference between the simulation and the theoretical prediction.}
    \label{fig:r}
\end{figure*}

We now aim to affirm and demonstrate the above results by simulating wave-packet dynamics numerically. To this end, we choose a one-dimensional non-Hermitian Dirac model with the Hamiltonian

\begin{equation}
    H_0=\sum_n (-\frac{i}{2}\ket{n}\bra{n+1} + H.c)\sigma_x  + m\ket{n}\bra{n}\sigma_z,
\end{equation}
where the mass $m$ is in general complex, and the lattice constant has been set to $1$. Such a model may be physically realized, for instance, by applying a staggered dissipation (gain) to a 1D chain of coupled oscillators, thereby creating a complex mass term. This is in complete analogy to how a real mass term may be realized by applying a staggered potential to the same chain \cite{neder2023bloch}.
The $k$-space representation of this model under periodic boundary conditions is
\begin{equation}
    H_0(k)=
    \begin{pmatrix}
    m & \sin k \\
    \sin k  & -m
    \end{pmatrix},
\end{equation}
and the energy spectrum is
\begin{equation}
    \varepsilon_{\pm,k}=\pm\sqrt{m^2+\sin^2 k}=\pm\varepsilon(k).
\end{equation}
The above model admits no band degeneracies unless both $\text{Re}(m)=0$ and $\vert\text{Im}(m)\vert\leq1$. Furthermore, the branches of $\varepsilon$ can be defined such that the imaginary part of the positive branch is guaranteed to always be non-negative while the imaginary part of the negative branch is guaranteed to always be non-positive. These properties mean that, for all choices of $m$ but those stated above, the positive branch remains the maximally amplified band across the entire Brillouin zone and the non-Hermitian adiabatic theorem may be safely applied to states confined to that band.

The state of the system is initialized as a single-band localized Gaussian wave-packet around some central momentum $k_c$ and central position $r_c$ far away from the edge,
\begin{equation}
    \ket{\psi_0(k_c,r_c)} = \int_k \frac{1}{(2\pi\sigma^2)^\frac{1}{4}}e^{-\frac{(k-k_c)^2}{4\sigma^2}}\ket{\psi^R_{+}(k)}e^{-ikr_c},
\end{equation}
where $\sigma$ is the $k$-space width of the wavepacket and the eigenfunctions of the amplified band of $H_0$ are given in the following gauge, which is valid for all values of $k$ when $m$ lies in the upper half of the complex plane,
\begin{equation}
    \label{eq:dirac_eigenfunc}
    \ket{\psi^R_{+}(k)} = \frac{e^{ikr}}{\sqrt{\vert\varepsilon(k)+m\vert^2+\sin^2 k }}
    \begin{pmatrix}
    \varepsilon(k)+m \\ \sin k    
    \end{pmatrix}.
\end{equation}
A complex electric field is added to the model by including an external potential
\begin{equation}
    V = \sum_{n}eEn\ket{n}\bra{n}.
\end{equation}
While this realization of the field does not adhere to periodic boundary conditions, there was no discernible change in the simulated dynamics between open and periodic boundary conditions, so long as the wavepacket remains well away from the edges. This is in accordance with the idea that the dynamics of a state localized away from the edges depend only on the properties of the bulk \cite{schomerus_fundamental_2022}. 
\begin{table}[t]
\caption{Model parameters used in numerical simulations. The chosen values are representative.}
\label{table:modelParams}
\begin{tabular}{p{0.3\linewidth} p{0.3\linewidth} p{0.3\linewidth}}
    \hline\hline
      & $m$ & $eE$ \\
     \hline
    (I) & $2$ & $0.001$\\
    (II) & $2+0.1i$ & $0.001$\\
    (III) & $2+10i$ & $0.001$\\
    \hline
    (IV) & $2$ & $0.001i$\\
    (V) & $2+0.1i$ & $0.001i$\\
    (VI) & $2+10i$ & $0.001i$\\
    \hline
    (VII) & $2$ & $0.001(1+i)$\\
    (VIII) & $2+0.1i$ & $0.001(1+i)$\\
    (IX) & $2+10i$ & $0.001(1+i)$\\
     \hline\hline
\end{tabular}
\end{table}
The same simulation process was repeated several times for different choices of model parameters $m$ and $eE$, as detailed in Table \ref{table:modelParams}. The different parameter choices cover Hermitian, weakly non-Hermitian, and strongly non-Hermitian base models subject to purely real, purely imaginary and complex fields.

As explained above, none of the detailed parameter choices are close to a degeneracy or to an exceptional point, and for all of them the positive branch \eqref{eq:dirac_eigenfunc} remains the maximally amplified band for all values of $k$. This remains unchanged by the addition of the field, which is chosen weak enough compared to $m$ to remain in the perturbative regime. The numerical values of the parameters chosen here are representative.

The numerical results for the simulated norm-growth and velocity for parameter choices (I)-(VI) are presented in figures \ref{fig:N}, \ref{fig:r}, respectively. As is shown if the figures, the results for the real field (parameters (I)-(III)) are qualitatively different from those of the purely imaginary field (parameters (IV)-(VI)). Conversely, results for the complex field (parameters (VII)-(IX)) have proven nearly identical to those for the purely imaginary field (parameters (IV)-(VI)). For the sake of completeness, they are presented in appendix \ref{App:graphs}.

The results are shown after subtraction of the field-independent contributions, $2\text{Im}\varepsilon(k_c)$  and $\partial_k\text{Re}\varepsilon(k_c)$, respectively. Additional corrections to the norm-growth and velocity arising due to the finite width of the wave-packet $\sigma$ were also calculated and subtracted from the simulated results (see Appendix \ref{App:width})
\begin{subequations}
    \label{eq:field_induced}
    \begin{align}
        \left(\frac{\dot{N}}{N}\right)_E & =\frac{\dot{N}}{N} - 2\text{Im}\varepsilon(k_c) - \text{\{finite-width terms\}}, \label{subeq:N_E}\\
        v_E & = \dot{r}_c - \partial_k\text{Re}\varepsilon(k_c) - \text{\{finite-width terms\}}. \label{subeq:v_E}
    \end{align}
\end{subequations}
The results of the simulations are compared to the theoretical predictions of \eqref{eq:EOM_N}, \eqref{eq:EOM_r}, across the different choices of model parameters. For all simulations, the initial central momentum was arbitrarily chosen to be $k_c=1$, which has no particular symmetry.

Since the localization in real space is proportional to $1/\sigma$, the $k$-space width is bounded from below by the scale of the inverse system size. A relatively large system size $L=1000$ was therefore chosen so as to allow for a small packet width $\sigma=0.0126$.
We note that, although the scale of the geometric effects on the norm-growth and velocity are relatively small in our chosen model, all parameter choices show agreement with the theoretical prediction to a very high precision, up to small decaying oscillations around the predicted value (see Appendix \ref{App:Simulations}).

Since we are particularly interested in the role of the quantum metric in the anomalous velocity, we also demonstrate explicitly the possibility of extracting the quantum metric from the velocity. To that end, we simulate the wave-packet dynamics for parameter choice (IV) (see Table \ref{table:modelParams}). This choice of parameters is preferable as the unperturbed Hamiltonian is Hermitian and the field is purely imaginary, simplifying the calculations.

The numerically calculated velocity is averaged over one period of oscillations (see Appendix \ref{App:Simulations}) and then subtracted by all contributions and finite-width corrections (see Appendix \ref{App:width}), except for the anomalous velocity. The obtained value is then divided by $2\text{Im}(eE)$ to yield the estimation for the quantum metric at initial momentum $k_c$.

\begin{figure}[ht!]
    \centering
    \includegraphics[width=0.9\linewidth]{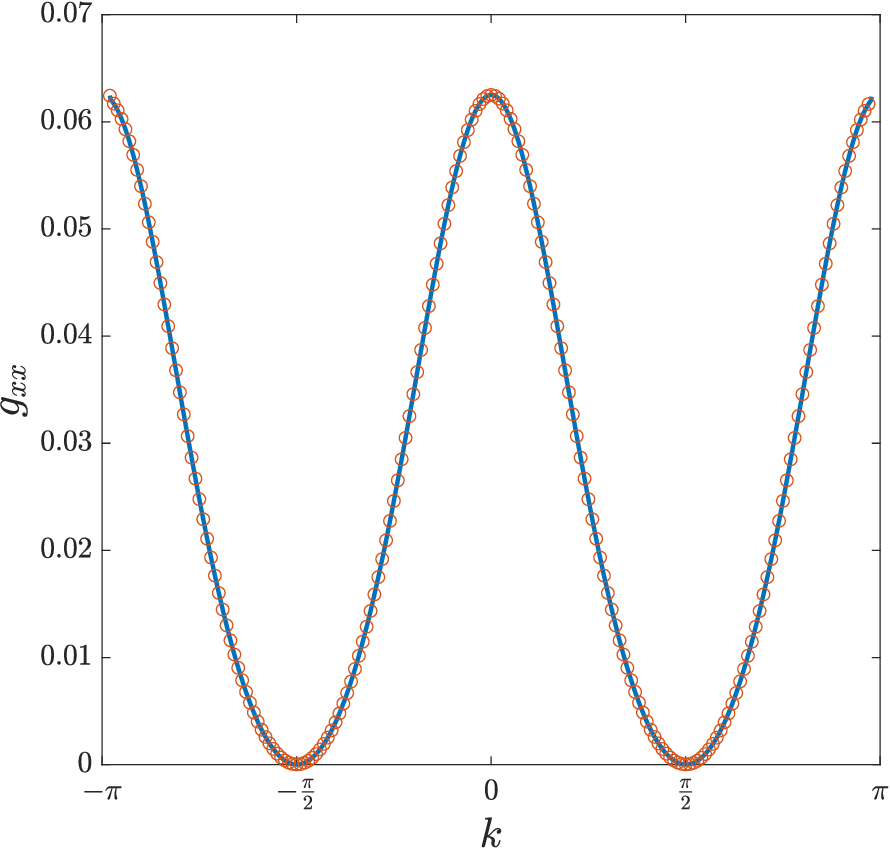}
    \caption{Comparison between the quantum metric calculated from the simulations using \eqref{eq:EOM_r_imag} (orange circles) and the theoretical one predicted using \eqref{eq:def_QuantumGeometricTensor},\eqref{eq:QGT_split} (solid blue line) for model (IV) from Table \ref{table:modelParams}.}.
    \label{fig:g}
\end{figure}

The process is repeated for many values of $k_c$ throughout the Brillouin zone. The obtained values are compared in figure \ref{fig:g} to the theoretical prediction for the quantum metric, which is based on \eqref{eq:def_QuantumGeometricTensor},\eqref{eq:QGT_split}, and \eqref{eq:dirac_eigenfunc}.

The agreement shown above in figures \ref{fig:N}, \ref{fig:r} between the simulated and theoretical results does not persist for arbitrarily long times. Specifically, after relatively long times (more than ten times longer than those shown in the figures), the simulated wavepacket undergoes a sudden and short `collapse', after which its dynamics differ significantly from the theoretical prediction. In all parameter choices we have tested, following the collapse, the simulated wavepacket was concentrated in $k$-space around values of $k$ for which $\text{Im}\varepsilon_k$ is maximal, regardless of its initial conditions. The duration of stability before the collapse was not universal, however, and depended on the particular parameter choices. We emphasize that the collapse occurs even in the complete absence of fields, and only for non-Hermitian band Hamiltonians (i.e. complex values of $m$). The duration of stability was found to also depend on the numerical precision of the calculations, with the duration growing longer with increased precision. We therefore hypothesize that the imperfect precision of numerical calculations, and the resulting rounding errors, may behave like an extremely weak effective disorder in the system, against which the non-Hermitian band Hamiltonian is not stable, due to the relative gain of modes with high $\text{Im}\varepsilon_k$ \cite{spring_phase_2023}. We leave a more thorough investigation of this phenomenon to future works.

\section{Summary and conclusions}\label{sec:Conclusions}
In this work, we derived the equations of motion for a non-Hermitian Bloch electron subject to a complex electric field. Our results reveal new anomalous terms which arise due to the complex nature of the field and are relevant even when the unperturbed system is perfectly Hermitian. Most importantly, we derived a `generalized anomalous velocity' which couples the field to the entirety of the Quantum Geometric Tensor, including the quantum metric. This quantum metric dependent velocity appears regardless of whether the unperturbed Hamiltonian is Hermitian or not. While the physical significance of the quantum metric has already been investigated elsewhere, its relevance has been mostly confined to higher order effects or strongly correlated systems \cite{LiangPhysRevB.96.064511,Gao_PhysRevLett.112.166601,bleu_effective_2018,Holder_PhysRevResearch.2.033100,leblanc_universal_2021,NeupertPhysRevB.87.245103,ozawa_steady-state_2018}. Our work reveals that it in fact arises naturally even in non-interacting systems and already to first order, provided non-Hermitian perturbations are considered. Additionally, a `gain velocity' term which couples the electric field to the real-space spread of the wave-packet was also uncovered.
Lastly, we generalized known anomalous terms involving the difference of Berry connections \cite{silberstein_berry_2020} and, using our formalism, unambiguously revealed them as expressing a first-order perturbative correction to the band energy arising from an interplay between the external field and the non-orthogonality of the eigenfunctions of the different bands.

In order to demonstrate our results, we conduced numerical simulations in a one-dimensional non-Hermitian lattice Dirac model. These simulations clearly show agreement with our equations of motion and demonstrate the possibility of extracting the quantum metric from the generalized anomalous velocity.

Our results in Section \ref{sec:EOM} further serve to highlight an often glossed-over peculiarity of non-Hermitian topological band theory. Namely, that while the non-Hermitian Berry phase which appears in the adiabatic theorem is given by a contour integral of $A^{LR}$, the geometric quantities which appear in the anomalous velocity \eqref{eq:EOM_r} are independent of the left eigenbasis. Specifically, the Berry curvature is given by the curl of $A^{RR}$, and not $A^{LR}$. This implies that Berry phases are no longer given by surface integrals of the (physically relevant) Berry curvature. Our formalism, however, suggests an interpretation of this apparent discrepancy by splitting the so-called non-Hermitian Berry phase into two separate contributions. The first contribution is an energy correction proportional to $A^{LR}-A^{RR}$, which is gauge invariant for any contour in parameter space. The second contribution would then be a geometric phase given by a contour integral of $A^{RR}$, which is gauge invariant for closed loops in parameter space. This contribution is indeed a phase (in the sense that it is always real and defined up to a multiple of $2\pi$), and is directly responsible for the anomalous velocity. It would be interesting to investigate whether this interpretation naturally extends to frameworks involving complex crystal momenta \cite{YaoShunyuPhysRevLett.121.086803,YokomizoPhysRevLett.123.066404} in a way that agrees with our results for complex fields.

The theoretical formalism presented in this paper is not limited to treatment of electric fields and may be applied to any perturbation, be it Hermitian or not, within the single band approximation. For instance, we expect that this approach may prove useful to the investigation of more complex electromagnetic, lattice deformation and spin-orbit coupling effects.
An additional possible extension of our work is the incorporation of boundaries into the semi-classical theory. While the problem of describing wavepacket dynamics along topological edges which may be non-Hermitian remains challenging, we believe that our methods, along with recent advances regarding bulk-edge correspondence in non-Hermitian systems \cite{RapoportPhysRevB.107.085117}, may serve to further research on this topic.

Finally, going beyond the single-band approximation, one has to generalize the approach to allow for degenerate or nearly degenerate bands and incorporate multi-band effects into the equations of motion. We believe that the projector-based approach presented here may prove useful in the development of such a theory, in analogy to the non-Abelian Berry connections of the Hermitian setting \cite{SHINDOU2005399,CulcerPhysRevB.72.085110}.

\section*{acknowledgments}
The authors thank J. Behrends and H. Schomerus for useful discussions. M.G. has been supported by the Israel Science Foundation (ISF) and the Directorate for Defense Research and Development (DDR\&D) grant No. 3427/21, by the ISF grant No. 1113/23, and by the US-Israel Binational Science Foundation (BSF) Grant No. 2020072. R.I. was supported by the Israeli Science Foundation under grant No. 1790/18 and U.S.-Israel Binational Science Foundation (BSF) Grant No. 2018226.

\appendix

\section{Derivations of the equations of motion} \label{AppEOM}
In this section we will go over the derivations which lead to equations \eqref{eq:EOM_N}, \eqref{eq:EOM_k} and \eqref{eq:EOM_r} in detail. As we will be mostly interested in calculating various expectation values involving the position operator, we begin by writing an expression for its matrix element for any two Bloch functions \cite{balian_relation_1989,silberstein_berry_2020}
\begin{equation}
\label{eq:r_matrixelement}
\begin{split}
    &\bra{\psi^{\alpha^\prime}_{n^\prime k^\prime}}\hat{r}\ket{\psi^\alpha_{nk}}=
    \bra{u^{\alpha^\prime}_{n^\prime k^\prime}}e^{-ik^\prime r}(-i)\partial_k(e^{ikr})\ket{u^\alpha_{nk}}\\
    =&i\delta(k-k^\prime)\braket{u^{\alpha^\prime}_{n^\prime k} | \partial_k u^\alpha_{nk}}
    -i\partial_k \left( \delta(k-k^\prime)\braket{u^{\alpha^\prime}_{n^\prime k} | u^\alpha_{nk}} \right),
\end{split}
\end{equation}
where $\alpha,\alpha^\prime \in \{R,L\}$.

In the calculations detailed in the following sections we simplify the notation by setting the norm of the wavepacket to $N=1$. This may be done without loss of generality as all final expressions only involve expectation values, which are normalized by definition.

\subsection{Relative norm-growth}
In order to derive equation \eqref{eq:EOM_N} from \eqref{eq:Wexp_N} we simply need to evaluate the expectation value $\braket{P_n(\hat{r}-r_c)}_W$. We begin by introducing the convenient notation
\begin{equation}
    \label{eq:AntiProjector}
    Q_n=1-P_n=\sum_{m\neq n}\int_k \ket{\psi^R_{mk}}\bra{\psi^L_{mk}}
\end{equation}
for the complement of the band projector $P_n$

By recalling that the state $\ket{W}$ is confined to the $n$th band, we conclude that it is annihilated by $Q_n$ when acted upon from the left
\begin{equation}
    Q_n\ket{W} = 0,
\end{equation}
while keeping in mind that the same does \emph{not} necessarily hold when $Q_n$ acts on $\bra{W}$ from the right, due to the possible non-orthogonality of the bands.

The expectation value may therefore be simplified as follows:
\begin{equation}
    \braket{P_n(\hat{r}-r_c)}_W=\braket{\hat{r}-r_c}_W - \braket{Q_n(\hat{r}-r_c)}_W
    = -\braket{Q_n\hat{r}}_W,
\end{equation}
where we used the definition of the central position $r_c = \braket{\hat{r}}_W$. All that remains is to evaluate the simplified form explicitly. We do this by using \eqref{eq:r_matrixelement} to calculate a slightly more general expression
\begin{widetext}
\begin{equation}
\begin{split}
    \label{eq:fQnr}
    \braket{f(\hat{k})Q_n\hat{r}}_W&=\sum_{m\neq n}\int_q\int_{kk^\prime} w_{nk^\prime}^\ast w_{nk} \braket{\psi^R_{nk^\prime} |f(\hat{k})| \psi^R_{mq}}\bra{\psi^L_{mq}}\hat{r}\ket{\psi^R_{nk}}\\
    &=\sum_{m\neq n}\int_q\int_{kk^\prime} w_{nk^\prime}^\ast w_{nk}
    f(q)\delta(k^\prime-q) \braket{u^R_{nk^\prime} | u^R_{mq}}i\delta(k-q)\braket{u^L_{mk} | \partial_k u^R_{nk}}\\
    &=\int_k \lvert w_{nk}\rvert^2f(k)
    i\bra{u^R_{nk}} \sum_{m\neq n}\ket{u^R_{mk}}\bra{u^L_{mk}} \ket{\partial_k u^R_{nk}}
    =\int_k \lvert w_{nk}\rvert^2 I_{nn}(k)f(k)(A^{RR}-A^{LR})\\
    &=f(k_c)(A^{RR}-A^{LR})\vert_{k=k_c},
\end{split}
\end{equation}
\end{widetext}
where we have made use of the properties of the Gram matrix, the assumption that the envelope function $g(k)$ is sharply peaked around $k_c$ and the resolution of identity $1=\sum_m \ket{u^R_{mk}}\bra{u^L_{mk}}$. By taking the case $f(k)=1$ we arrive at
\begin{equation}
    \label{eq:Qnr}
    \braket{Q_n\hat{r}} = (A^{RR}-A^{LR})\vert_{k=k_c}.
\end{equation}
Plugging this expression into \eqref{eq:Wexp_N} indeed yields the result quoted in the main text \eqref{eq:EOM_N}.

\medskip
\medskip

\subsection{momentum}
We begin by writing equation \eqref{eq:SB_Ehrenfest} under constant electric field \eqref{eq:ElecPotential} for an observable which may be any arbitrary function of the momentum $\hat{O} = f(\hat{k})$
\begin{equation}
    \label{eq:Ehrenfest_k}
    \begin{split}
    \frac{d}{dt}\braket{f(\hat{k})}_W =&
    2\text{Im}\left[ \braket{f(\hat{k})H_0}_W - \braket{f(\hat{k})}_W\braket{H_0}_W \right. \\
    &\left. +eE \left( \braket{f(\hat{k})P_nr}_W - \braket{f(\hat{k})}_W\braket{P_nr}_W \right)\right].       
    \end{split}
\end{equation}

This expression may be simplified by noticing that since both $f(\hat{k})$ and $H_0$ are diagonal in the Bloch basis and that the packet is sharply centered around some $k_c$, one may immediately conclude that
\begin{equation}
    \braket{f(\hat{k})H_0}_W = (f(k)\varepsilon_{nk})\vert_{k=k_c},
\end{equation}
and hence the first two terms of \eqref{eq:Ehrenfest_k} cancel. This is unsurprising as these terms are nothing but the equation of motion in the absence of fields. For the other terms, a more direct calculation is required
\begin{widetext}
\begin{equation}
\begin{split}
    \braket{f(\hat{k})P_nr}_W=&
    \int_{kk^\prime}\int_q w_{nk^\prime}^\ast w_{nk} \bra{\psi^R_{nk^\prime}}f(\hat{k})\ket{\psi^R_{nq}}  \bra{\psi^L_{nq}}\hat{r}\ket{\psi^R_{nk}}\\
    =&\int_{kk^\prime}w_{nk^\prime}^\ast w_{nk} I_{nn}(k^\prime)f(k^\prime)
    \left[ i\delta(k-k^\prime)\braket{u^{L}_{nk} | \partial_k u^R_{nk}}
    -i\partial_k \left( \delta(k-k^\prime)\braket{u^L_{nk} | u^R_{nk}} \right)\right]\\
    =&\int_{k}\lvert w_{nk}\rvert^2 I_{nn}(k)f(k)A^{LR}
    +i\int_{k}w_{nk}^\ast \partial_k w_{nk} I_{nn}(k)f(k),
\end{split}
\end{equation}
where we have again used the properties of the Gram matrix and equation \eqref{eq:r_matrixelement}. The first of these two final integrals evaluates simply to
\begin{equation}
    \int_{k}g(k)f(k)A^{LR} = A^{LR}(k_c)f(k_c).
\end{equation}
We proceed by introducing the notations $w_{nk}=\lvert w_{nk} \rvert e^{i\varphi(k)}$ and $z(k)=e^{i\varphi(k)}/\sqrt{I_{nn}(k)}$. This allows us, using \eqref{eq:def_envelope}, to substitute $w_{nk}=z(k)\sqrt{g(k)}$ in order to more easily express the second integral as
\begin{equation}
\begin{split} \label{eq:integral_1stmoment}
    i\int_{k}w_{nk}^\ast \partial_k w_{nk} I_{nn}(k)f(k)=&
    i\int_{k}\frac{\sqrt{g(k)}}{z(k)}\partial_k(z(k)\sqrt{g(k)}) f(k)\\
    =&i\int_{k}g(k)\frac{z^\prime(k)}{z(k)}f(k) +i\int_k\partial_k\left( \frac{g(k)}{2}\right)f(k)\\
    =& \alpha(k_c)f(k_c) - \frac{i}{2}f^\prime(k_c).
\end{split}
\end{equation}

\end{widetext}
In these evaluations we have made use of the fact that the envelope function $g(k)$ is sharply peaked around $k_c$ and have also introduced the notation $\alpha(k)=iz^\prime(k)/z(k)$ for brevity. Combining the above results yields
\begin{equation}
    \label{eq:integral_kPr_final}
    \braket{f(\hat{k})P_nr}_W= (A^{LR}(k_c)+\alpha(k_c))f(k_c) - \frac{i}{2}f^\prime(k_c).
\end{equation}

By setting $f(\hat{k})=1$ we immediately conclude that
\begin{equation}
    \label{eq:Pnr}
    \braket{P_nr}_W = A^{LR}(k_c)+\alpha(k_c),
\end{equation}
and therefore we may write in general
\begin{equation}
    \braket{f(\hat{k})P_nr}_W=\braket{f(\hat{k})}_W\braket{P_nr}_W - \frac{i}{2}f^\prime(k_c).
\end{equation}
Applying this to \eqref{eq:Ehrenfest_k} with $f(\hat{k})=\hat{k}$, we arrive at
\begin{equation}
    \frac{d}{dt}\braket{f(\hat{k})}_W = 2\text{Im}\left( -\frac{i}{2}eE \right) = -\text{Re}(eE),
\end{equation}
as quoted in the main text.

\medskip
\bigskip

\subsection{velocity}
In this section, we must take care to differentiate between the coordinate of the observable $\hat{O}=\hat{r}_i$ and the one coupled to the external electric field: $E_j\hat{r}_j$. Making use of \eqref{eq:AntiProjector}, we may write equation \eqref{eq:SB_Ehrenfest} for the position of the wave-packet as follows:
\begin{equation}
    \begin{split}
    \label{eq:velocity_Expvals}
    \frac{d}{dt}\braket{\hat{r}_i}_W =
    2\text{Im}&\left[ \braket{\hat{r}_iH_0}_W - \braket{\hat{r}_i}_W\braket{H_0}_W \right. \\
    &+eE_j\left( \braket{\hat{r}_i\hat{r}_j}_W - \braket{\hat{r}_i}_W\braket{\hat{r}_j}_W\right) \\
    & \left. -eE_j\left( \braket{\hat{r}_iQ_n\hat{r}_j}_W - \braket{\hat{r}_i}_W\braket{Q_n\hat{r}_j}_W \right) \right].
    \end{split}
\end{equation}

The first line of the above expression is simply the velocity in the absence of fields which is known to be \cite{schomerus_non-hermitian-transport_2014,xu_weyl_2017,silberstein_berry_2020} $\partial_{k_i}\text{Re}(\varepsilon_{nk})$. The terms on the second line plainly form the covariance induced velocity $2\text{Im}(eE_j)\Sigma_{ij}$ which appears in the main text. We are now left with the task of evaluating the terms on the third and final line.

The expectation value of the position may be readily calculated by making use of equations \eqref{eq:Qnr} and \eqref{eq:Pnr}
\begin{equation}
    \label{eq:r}
    \braket{\hat{r}_i}_W = \braket{P_n\hat{r}_i}_W + \braket{Q_n\hat{r}_i}_W = A^{RR}_i(k_c)+\alpha_i(k_c).
\end{equation}
We are left only with the somewhat lengthy yet straightforward calculation of the expectation value $\braket{\hat{r}_iQ_n\hat{r}_j}_W$. We introduce the abbreviated notation $\partial_i$ for $\frac{\partial}{\partial k_i}$ and $A^{LR}_i, A^{RR}_i$ for the Berry connections with respect to the parameter $k_i$.

\begin{widetext}

\begin{equation}\label{eq:rQr}
\begin{split}
    \braket{\hat{r}_iQ_n\hat{r}_j}_W=&
    \sum_{m\neq n} \int_{kk^\prime}\int_q w_{nk^\prime}^\ast w_{nk}
    \bra{\psi^R_{nk^\prime}}\hat{r}_i\ket{\psi^R_{mq}}
    \bra{\psi^L_{mq}}\hat{r}_j\ket{\psi^R_{nk}}\\
    =&\sum_{m\neq n} \int_{kk^\prime}\int_q w_{nk^\prime}^\ast w_{nk}
    \bra{\psi^R_{nk^\prime}}\hat{r}_i\ket{\psi^R_{mq}}i\delta(k-q)\braket{u^L_{mk}|\partial_j u^R_{nk}}\\
    =&\sum_{m\neq n} \int_{kk^\prime} w_{nk^\prime}^\ast w_{nk}
    \left(  i\delta(k-k^\prime)\braket{u^R_{nk} | \partial_i u^R_{mk}}
    -i\partial_i \left( \delta(k-k^\prime)\braket{u^R_{nk} | u^R_{mk}} \right)  \right)
    i\braket{u^L_{mk}|\partial_j u^R_{nk}}\\
    =& -\sum_{m\neq n}\int_{k} \lvert w_{nk}\rvert^2 
    \left( \braket{u^R_{nk} | \partial_i u^R_{mk}} \braket{u^L_{mk}|\partial_j u^R_{nk}}
    +\braket{u^R_{nk}|u^R_{mk}}\braket{\partial_i u^L_{mk}|\partial_j u^R_{nk}} \right) \\
    &-\sum_{m\neq n}\int_{k} \lvert w_{nk}\rvert^2 \braket{u^R_{nk}|u^R_{mk}}\braket{ u^L_{mk}|\partial_i\partial_j u^R_{nk}}
    -\sum_{m\neq n} \int_{k} w_{nk}^\ast \partial_i w_{nk}
    \braket{u^R_{nk}|u^R_{mk}}\braket{ u^L_{mk}|\partial_j u^R_{nk}}\\
    =& -\int_{k} \lvert w_{nk}\rvert^2 
    \bra{u^R_{nk}}\partial_i \left( \sum_{m\neq n}\ket{u^R_{mk}}\bra{u^L_{mk}} \right)
    \ket{\partial_j u^R_{nk}} \\
    &- \int_{k} \lvert w_{nk}\rvert^2 \bra{u^R_{nk}}
    \left( \sum_{m\neq n}\ket{u^R_{mk}}\bra{u^L_{mk}} \right)\ket{ \partial_i\partial_j u^R_{nk}}
    - \int_{k} w_{nk}^\ast \partial_i w_{nk}
    \bra{u^R_{nk}}\left( \sum_{m\neq n}\ket{u^R_{mk}}\bra{u^L_{mk}} \right)\ket{\partial_j u^R_{nk}}\\
    =& \int_{k} \lvert w_{nk}\rvert^2 
    \left( \braket{u^R_{nk}|\partial_i u^R_{nk}} \braket{u^L_{nk}|\partial_j u^R_{nk}}
    +  \braket{u^R_{nk}|u^R_{nk}} \braket{\partial_i u^L_{nk}|\partial_j u^R_{nk}} + \braket{u^R_{nk}|u^R_{nk}}\braket{u^L_{nk}|\partial_i\partial_j u^R_{nk}} \right)\\
    &+ \int_{k} \lvert w_{nk}\rvert^2
    \left( - \braket{u^R_{nk}|\partial_i\partial_j u^R_{nk}}\right)\\
    &+ \int_{k} w_{nk}^\ast \partial_i w_{nk}
    \left( \braket{u^R_{nk}|u^R_{nk}}\braket{u^L_{nk}|\partial_j u^R_{nk}} - \braket{u^R_{nk}|\partial_j u^R_{nk}} \right).
\end{split}
\end{equation}
We term the last three integrals $I_1,I_2,I_3$ for convenience and evaluate each in turn. The first integral is simply:
\begin{equation}
    \begin{split}
        I_1 = \int_k g(k) \left( -A^{RR}_iA^{LR}_j-i\partial_i A^{LR}_j \right)
        =\left( -A^{RR}_iA^{LR}_j -i\partial_i A^{LR}_j\right)(k_c).
    \end{split}
\end{equation}
The second is somewhat more involved
\begin{equation}
    \begin{split}
        I_2 =& \int_k g(k)
        \left( \frac{\braket{\partial_i u^R_{nk} | \partial_j u^R_{nk}}}{I_{nn}(k)} - \frac{\partial_i \braket{u^R_{nk}|\partial_j u^R_{nk}}}{I_{nn}(k)} \right)\\
        =& \int_k g(k)
        \left( \frac{\braket{\partial_i u^R_{nk} | \partial_j u^R_{nk}}}{I_{nn}(k)} - \frac{\partial_i I_{nn}(k) \braket{u^R_{nk}|\partial_j u^R_{nk}}}{I^2_{nn}(k)} 
        +i \partial_i A^{RR}_j\right)\\
        =& \int_k g(k)
        \left( \frac{\braket{\partial_i u^R_{nk} | \partial_j u^R_{nk}}}{I_{nn}(k)} - \frac{\braket{\partial_i u^R_{nk}| u^R_{nk}} \braket{u^R_{nk}|\partial_j u^R_{nk}}}{I^2_{nn}(k)} 
        + A^{RR}_i A^{RR}_j+i \partial_i A^{RR}_j\right)\\
        =& \left(T_{ij} + A^{RR}_i A^{RR}_j +i\partial_i A^{RR}_j \right)(k_c),
    \end{split}
\end{equation}
where we identified the definition of the quantum geometric tensor $T_{ij}$ \eqref{eq:def_QuantumGeometricTensor}. The last integral may be evaluated using \eqref{eq:integral_1stmoment}
\begin{equation}
    -i\int_{k} w_{nk}^\ast \partial_i w_{nk} I_{nn}(k)
    \left( A^{LR}_j - A^{RR}_j \right) = -\alpha_i(k_c)\left( A^{LR}_j - A^{RR}_j\right)(k_c) +\frac{i}{2}\partial_i\left( A^{LR}_j - A^{RR}_j\right)\vert_{k=k_c}.
\end{equation}
Combining all these results and making use of \eqref{eq:Qnr},\eqref{eq:r} we finally arrive at
\begin{equation}
    \begin{split}
    \label{eq:rQr_final}
    \braket{\hat{r}_iQ_n\hat{r}_j}_W=&
    T_{ij}
    -\frac{i}{2}\partial_i\left( A^{LR}_j - A^{RR}_j\right)
    -\left( A^{RR}_i +\alpha_i\right)\left( A^{LR}_j - A^{RR}_j \right)\\
    =& \braket{\hat{r}_i}_W\braket{Q_n\hat{r}_j}_W
    + T_{ij} -\frac{i}{2}\partial_i\left( A^{LR}_j - A^{RR}_j\right).
    \end{split}
\end{equation}
\end{widetext}
Plugging this result into \eqref{eq:velocity_Expvals} finally yields
\begin{equation}
\begin{split}
    \dot{r}_i =&\partial_i \text{Re}\left[ \varepsilon_{nk} + eE\cdot \left(A^{LR} - A^{RR}\right) \right]\\
    & -2\text{Im}\left(eE_jT_{ij}\right) + 2\text{Im}(eE_j)\Sigma_{ij},
\end{split}
\end{equation}
which is the result quoted in the main text as \eqref{eq:EOM_r}.

\section{Finite-width Corrections} \label{App:width}
As mentioned in previous sections, the theoretical calculations presented in this work rely repeatedly on the assumption that the wave packet is sharply peaked around the central momentum $k_c$, as per \eqref{eq:peakCondition}. However, any wave-packet which is also localized in real space must posses some finite characteristic $k$-space width $\sigma$, defined by
\begin{equation}
    \int_k g(k)(k-k_c)^2 = N\sigma^2.
\end{equation}

In the limit of small $\sigma$, and assuming higher cumulants of $g(k)$ are negligible, the corrected estimate for integrals modulated by the envelope function becomes
\begin{equation}
    \label{eq:widthCorrection_generic}
    \frac{1}{N}\int_k g(k)f(k) \approx f(k_c) + \frac{\sigma^2}{2}\partial^2_k f(k)\vert_{k=k_c}.
\end{equation}
This expression may be readily obtained by taking the Taylor expansion of $f(k)$ about $k_c$.

The above corrected expression produces small, width-dependent corrections to the equations of motion, even in the absence of external fields. Corrections which involve both the finite width and the external field of course also exist. They are, however, much smaller in comparison, since only weak fields and sharply-peaked packets are considered.

The corrections in the absence of fields may be calculated from \eqref{eq:Ndot_generic} \eqref{eq:SB_Ehrenfest} while taking $H=H_0$, i.e. $V=0$. Previous works have already shown the corrections to the norm-growth and momentum to be \cite{graefe_wave-packet_2011,silberstein_berry_2020}
\begin{equation}
    \label{eq:widthCorrection_Ndot}
    \frac{\dot{N}}{N} \approx \left( 2\text{Im}\varepsilon_{k} + \sigma^2\partial^2_k\text{Im}\varepsilon_k \right)_{k=k_c},
\end{equation}
\begin{equation}
    \label{eq:widthCorrection_k}
    \dot{k} \approx 2\sigma^2\partial_k\text{Im}\varepsilon_k\vert_{k=k_c}.
\end{equation}
It remains to calculate the correction for the velocity
\begin{equation}
    \dot{r} = 2\text{Im}\left( \braket{\hat{r}H_0}_W - \braket{\hat{r}}_W\braket{H_0}_W\right).
\end{equation}

When restricted to the $n$th band by acting on $\ket{W}$, the unperturbed Hamiltonian $H_0$ can be regarded as a function of the crystal momentum $\varepsilon_n(\hat{k})$. We may therefore re-write the first expectation value in the above equation as
\begin{equation}
\begin{split}
        \braket{\hat{r}\varepsilon_n(\hat{k})}_W &= \braket{\varepsilon_n(\hat{k})\hat{r}}_W + \braket{[\hat{r},\varepsilon_n(\hat{k})]}_W \\
        &=\braket{\varepsilon_n(\hat{k})P_n\hat{r}}_W + \braket{\varepsilon_n(\hat{k})Q_n\hat{r}}_W + i\braket{\partial_k\varepsilon_n(\hat{k})}_W
\end{split}
\end{equation}

By considering the expressions leading up to equations \eqref{eq:fQnr}, \eqref{eq:integral_kPr_final}, we may write the expectation value in integral form and apply the approximation \eqref{eq:widthCorrection_generic}
\begin{equation}
\begin{split}
    \braket{\hat{r}\varepsilon_n(\hat{k})}_W =& \int_k g(k)
    \left[ (A^{RR}+\alpha)(k)\varepsilon_n(k) + \frac{i}{2}\partial_k\varepsilon_n(k) \right] \\
    =& \beta(k_c)\varepsilon_n(k_c) + \frac{i}{2}\partial_k\varepsilon_n(k_c) \\
    & + \frac{\sigma^2}{2}\partial^2_k
    \left( \beta\varepsilon_n + \frac{i}{2}\partial_k\varepsilon_n\right)_{k=k_c},
\end{split}
\end{equation}
where we have introduced the abbreviated notation $$\beta(k) = A^{RR}(k) + \alpha(k) = \text{Re}A^{RR}(k)-\varphi^\prime(k).$$

By applying the same approximation to $\braket{\hat{r}}_W$, $\braket{H_0}_W$, and keeping only terms up to order $\sigma^2$, we arrive at the corrected expression for the unperturbed velocity
\begin{equation}
    \dot{r} \approx \left( \partial_k\text{Re}\varepsilon_n + \frac{\sigma^2}{2}\partial_k^3\text{Re}\varepsilon_n +2\sigma^2\partial_k\text{Im}\varepsilon_n\partial_k\beta\right)_{k=k_c}.
\end{equation}

The first two terms are the corrected mean value of $\partial_k\text{Re}\varepsilon_n$ felt across a packet of width $\sigma$. The last term is proportional to the drift in $k$ space \eqref{eq:widthCorrection_k} and exists only in non-Hermitian systems.

The function $\beta(k)$ is real-valued, and depends on the wave packet through the phase of the envelope $\varphi(k)$. By \eqref{eq:r}, it also has the property $\beta(k_c)=r_c$. We note that the phase of the envelope also develops with time as
\begin{equation}
    \varphi(k,t) = \varphi(k,0) - \text{Re}\varepsilon_n(k)t.
\end{equation}
This term expresses slight deformation of the wavepacket and results in increased (decreased) velocity exactly when the high (low) velocity components of the packet are more amplified than others.

\section{Anadiabatic oscillations} \label{App:Simulations}
Throughout the simulations shown in section \ref{sec:numerics}, rapid oscillations of the dynamical variables about their mean values can be seen. These oscillations are most visible in the velocity, and for Dirac models with real mass $m$ (parameters (I), (IV) and (VII) in Table \ref{table:modelParams}).

\begin{figure*}[t]
    \centering
     \includegraphics[width=\textwidth]{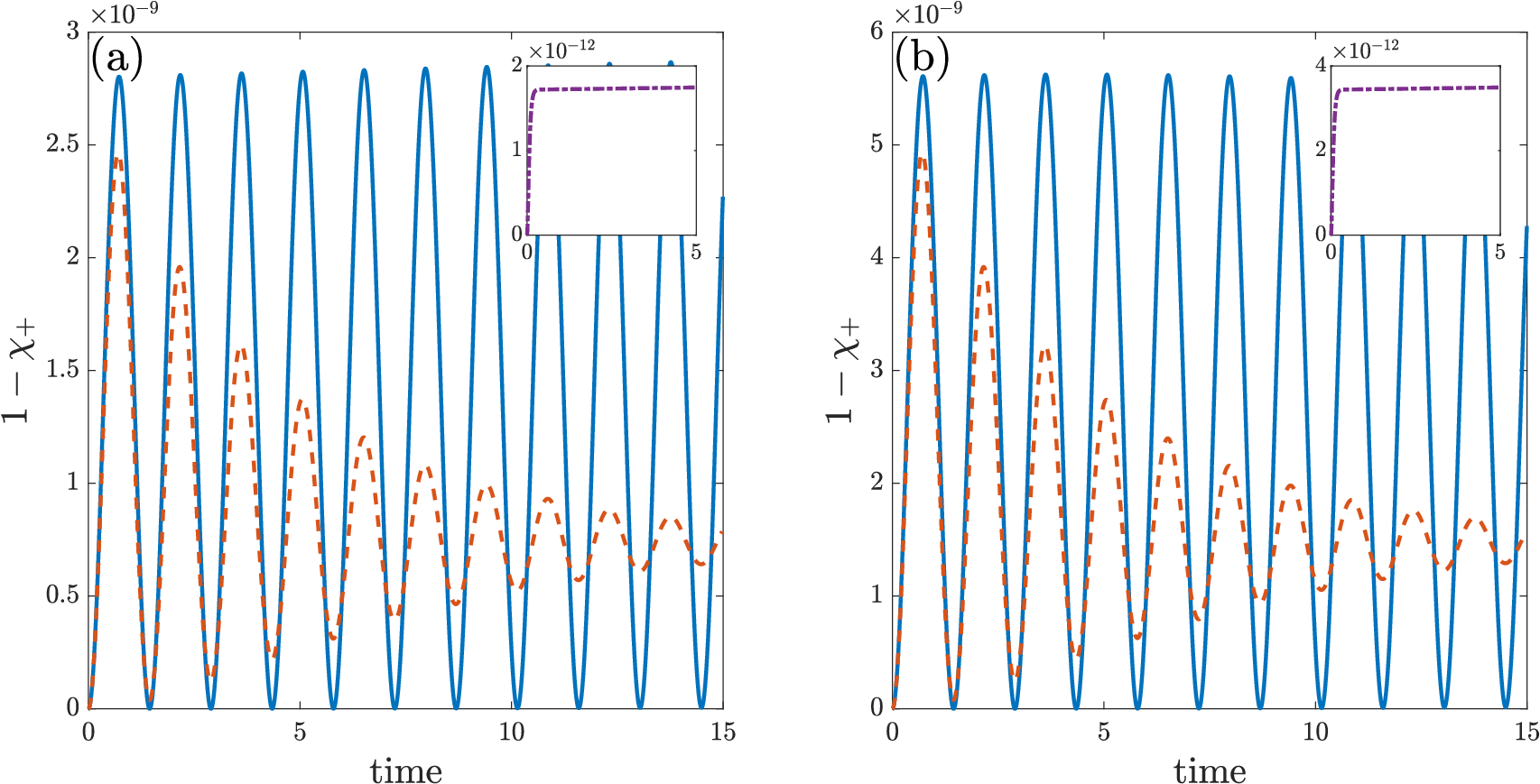}
    \hfill
    \caption{Time evolution of the deviation of the normalized single-band norm $\chi_+$ from the ideal value $1$ for the model parameters summarized in Table \ref{table:modelParams}. (a) Parameter choices (I)-(III) shown by solid blue line, dashed orange line, and dash-dotted purple line, respectively, the last of which is shown in the inset. (b) Same for parameter choices (VII)-(IX). Parameter choices (IV)-(VI) (not shown) yield nearly identical results to choices (I)-(III), respectively. Time is shown in units where $\hbar=1$. Simulations were conducted with $L=1000$ sites, packet-width $\sigma=0.0126$, and initial momentum $k_c=1$. The timestep used for each simulation was $0.01/\vert m\vert$.}
    \label{fig:bandNorm}
\end{figure*}

In order to investigate the origin of these oscillations, we introduce the normalized single-band norm:
\begin{equation}
    \label{eq:Singleband_N}
    \chi_n = \frac{N_n}{\sum_i N_i},
\end{equation}
where $N_i = \lVert P_i\ket{\psi}\rVert^2$ are the single-band norms. We note that, for non-orthogonal bands, the single band norms are \emph{not} necessarily additive. i.e. $N_i + N_j \neq \lVert (P_i+P_j)\ket{\psi} \rVert^2$.

In the ideal limit where the adiabatic theorem holds exactly, a packet initialized in the maximally amplified band should remain in that band throughout the evolution: $\chi_n(t)=1$. However, since any realistic evolution occurs at some finite rate, some anadiabaticity is inevitably introduced.

As is shown in figure \ref{fig:bandNorm}, in the course of our simulations $\chi_+$ dips from the ideal value of $1$ in an oscillatory fashion. The frequency of these oscillations was found to match those observed in the dynamical variables and be equal to the real energy gap at the wave-packet center
\begin{equation}
    \label{eq:ZBW_period}
    \omega = 2\text{Re}\varepsilon(k_c).
\end{equation}
We are therefore led to conclude that these oscillations originate from an anadiabatic correction to the dynamics, caused by rapid interference with states from the suppressed band.

This Zitterbewegung-like effect is already well known in Hermitian systems \cite{jiang_semiclassical_2005,zawadzki_zitterbewegung_2011}, and, for a two-band system (such as the Dirac model used in our simulations), is often understood as the rapid precession of the pseudo-spin vector $\mathbf{S}(t)$, representing the state of the system on the Bloch sphere, about the vector $\mathbf{d}(k(t))$, representing the (Hermitian) two-level Bloch Hamiltonian $H(k)=\mathbf{d}(k)\cdot\mathbf{\sigma}$, as $\mathbf{d}$ evolves in time due to changes in $k$ induced by the electric field.

When non-Hermiticity is introduced into the model through a complex value of $m$, the observed oscillations are seen to be damped over time, with the degree of damping increasing with the imaginary component of $m$. Quantitatively, the rate of damping in the oscillations of $\chi_+$ seem to be well approximated by
\begin{equation}
    \gamma \approx 2\text{Im}\varepsilon(k_c).
\end{equation}
This approximation was observed to be much less accurate for the oscillations in the velocity, however, although still within the correct order of magnitude.
As can be seen in figure \ref{fig:bandNorm}, through the damping, $\chi_+$ converges with time to some value lower than $1$, with the difference being seemingly proportional to $\lvert E\rvert^2$. While it is perhaps unsurprising that the introduction of non-Hermiticity, which is related to gain and loss, creates a damping effect on the precessory oscillations, the method by which such an effect occurs, as well as its quantitative description, are not immediately clear.

\begin{figure}[ht]
    \centering
    \includegraphics[width=0.9\linewidth]{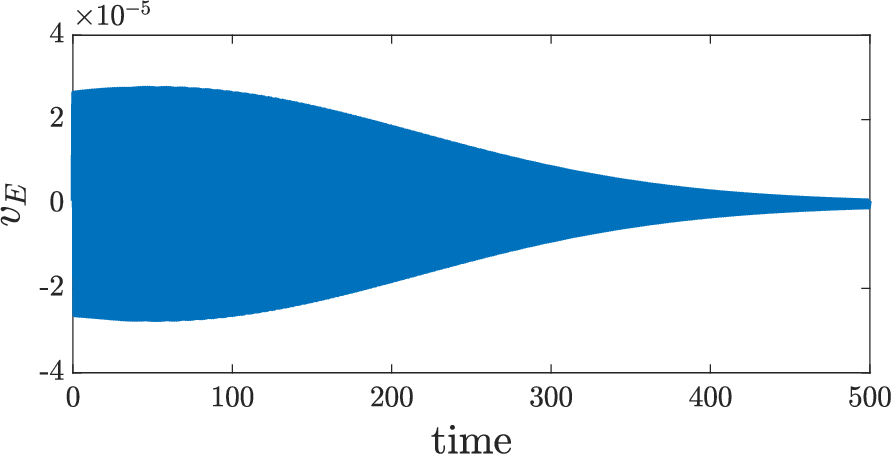}
    \caption{Decay of the anadiabatic oscillations of the velocity in a Hermitian model (parameter choice (I) in table \ref{table:modelParams}) over a long time. Simulation parameters identical to those used in figure \ref{fig:r}.}.
    \label{fig:r_long}
\end{figure}

When $m$ is taken to be real, the anadiabatic oscillations remain transient, albeit with a much longer decay period (see figure \ref{fig:r_long}). The transiency of Zitterbewegung in Hermitian condensed matter systems has been discussed before \cite{zawadzki_zitterbewegung_2011} and general arguments have shown it to be a result of the finite width of the wave-packet in $k$-space \cite{Lock10.1119/1.11697}.

Since for real values of $m$ the oscillations in the velocity decay very slowly compared to the oscillation period, their initial amplitude may be readily estimated. The initial amplitude was found to be numerically close to the value
\begin{equation}
    \label{eq:ZBW_amplitude}
    v_{ZB} \approx 2\lvert E\rvert g_{xx}(k_c).
\end{equation}
The above estimation was tested for several different values of $k_c$, (real) $m$ and $E$ and was found to be consistent. For complex values of $m$, however, the rapid decay of the oscillations prevent a reasonably accurate estimation of the initial amplitude. While we did not find a similar hypothesized expression for the amplitude of the oscillations in $\chi_+$, it was found to be seemingly proportional to $\lvert E\rvert^2$.

The connection between the quantum metric and anadiabatic effects, at least in Hermitian systems, has been noted before \cite{Gao_PhysRevLett.112.166601,bleu_effective_2018,Holder_PhysRevResearch.2.033100,leblanc_universal_2021}. Perhaps more surprising is the apparent dependence of \eqref{eq:ZBW_amplitude} on the absolute value of the field, as it puts the real and imaginary components of $E$ on equal footing. This is in contrast to the precession description which relies on the time dependence of $k$, which in turn depends only on the real part of the field.

We believe the observations detailed above raise new and interesting questions about the interplay between anadiabatic and Zitterbewegung-like effects and the non-Hermitian properties of perturbed systems, including ones regarding the precision and generality of these observations in models beyond the one tested here. We leave further investigation of these questions to future works.

\begin{figure*}[t]
    \centering
    \includegraphics[width=\textwidth]{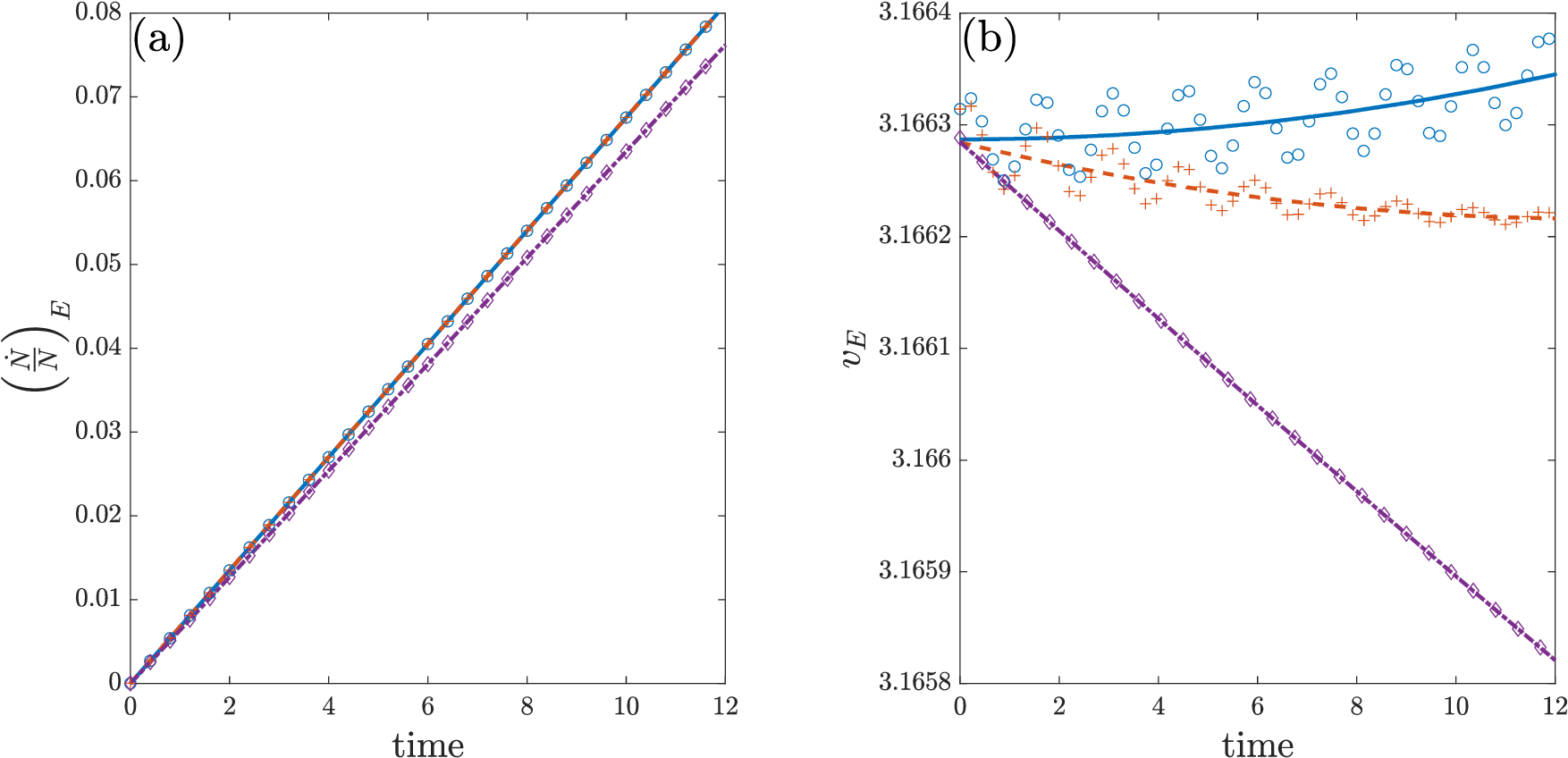}
    \hfill
    \caption{(a) Time evolution of the field-induced norm growth \eqref{subeq:N_E}, for the model parameters (VII)-(IX), detailed in Table \ref{table:modelParams}. Simulated results shown in blue circles, orange plus-signs, and purple diamonds, respectively. Theoretical predictions given by \eqref{eq:EOM_N} shown in solid blue line, dashed orange line, and dash-dotted purple line, respectively. (b) Time evolution of the field-induced velocity \eqref{subeq:v_E}, for the model parameters (VII)-(IX), detailed in Table \ref{table:modelParams}. Simulated results shown in blue circles, orange plus-signs, and purple diamonds, respectively. Theoretical predictions given by \eqref{eq:EOM_r} shown in solid blue line, dashed orange line, and dash-dotted purple line, respectively. Time is shown in units where $\hbar=1$. Simulations were conducted with $L=1000$ sites, packet-width $\sigma=0.0126$, and initial momentum $k_c=1$. The timestep used for each simulation was $0.01/\vert m\vert$.}
    \label{fig:complexfield}
\end{figure*}

\section{Relation between real-space localization and the quantum metric} \label{App:Cov}
In accordance with previous works which demonstrated a strong relation between the quantum metric and the real-space localization properties of wavefunctions in a band system \cite{marzari_maximally-localized_1997}, we aim to investigate the difference $\overline{\Sigma}$, as defined in \eqref{eq:def_CovReduced}, between the real-space covariance matrix and the quantum metric. To this end, we provide here an explicit calculation of the covariance $\Sigma_{ij}$.

Our main step is the calculation of the second moment of the position
\begin{equation}
    \label{eq:2ndMomentPosition}
    \braket{\hat{r}_i\hat{r}_j}_W = \braket{\hat{r}_iQ_n\hat{r}_j}_W + \braket{\hat{r}_iP_n\hat{r}_j}_W.
\end{equation}
Since the first term was already calculated in \eqref{eq:rQr_final}, it remains to similarly calculate the second one.

\begin{widetext}

\begin{equation} \label{eq:rPr}
\begin{split}
    \braket{\hat{r}_iP_n\hat{r}_j}_W=&
    \int_{kk^\prime}\int_q w_{nk^\prime}^\ast w_{nk}
    \bra{\psi^R_{nk^\prime}}\hat{r}_i\ket{\psi^R_{nq}}
    \bra{\psi^L_{nq}}\hat{r}_j\ket{\psi^R_{nk}}\\
    =&\int_{kk^\prime}\int_q w_{nk^\prime}^\ast w_{nk}
    \bra{\psi^R_{nk^\prime}}\hat{r}_i\ket{\psi^R_{nq}}
    (i\delta(k-q)\braket{u^L_{nk} | \partial_j u^R_{nk}}
    -i\partial_j\delta(k-q))\\
    =&\int_{kk^\prime}w_{nk^\prime}^\ast w_{nk}
    \bra{\psi^R_{nk^\prime}}\hat{r}_i\ket{\psi^R_{nk}}
    A^{LR}_j(k)
    +i\int_{kk^\prime} w_{nk^\prime}^\ast \partial_j w_{nk}
    \bra{\psi^R_{nk^\prime}}\hat{r}_i\ket{\psi^R_{nk}}\\
    =&\int_{kk^\prime}w_{nk^\prime}^\ast w_{nk}
    (i\delta(k-k^\prime)\braket{u^R_{nk} | \partial_i u^R_{nk}}
    -i\partial_i \left( \delta(k-k^\prime)\braket{u^R_{nk} | u^R_{nk}} \right))
    A^{LR}_j(k)\\
    &+i\int_{kk^\prime} w_{nk^\prime}^\ast \partial_j w_{nk}
    (i\delta(k-k^\prime)\braket{u^R_{nk} | \partial_i u^R_{nk}}
    -i\partial_i \left( \delta(k-k^\prime)\braket{u^R_{nk} | u^R_{nk}} \right))\\
    =&\int_{k}g(k)A^{RR}_i(k) A^{LR}_j(k)
    +i\int_{k}w_{nk}^\ast \partial_i w_{nk}I_{nn}(k)A^{LR}_j(k)
    +i\int_{k}g(k)\partial_i A^{LR}_j(k)\\
    &+i\int_{k} w_{nk}^\ast \partial_j w_{nk}I_{nn}(k)A^{RR}_i(k)
    -\int_{k} w_{nk}^\ast \partial_i\partial_j w_{nk}I_{nn}(k)\\
    =& \left( A^{RR}_i A^{LR}_j
    +\alpha_iA^{LR}_j + \alpha_jA^{RR}_i
    + \frac{i}{2}\partial_i A^{LR}_j - \frac{i}{2}\partial_j A^{RR}_i \right)_{k=k_c}
    -\int_{k} w_{nk}^\ast \partial_i\partial_j w_{nk}I_{nn}(k),
\end{split}
\end{equation}
where we have once more used \eqref{eq:r_matrixelement}, \eqref{eq:integral_1stmoment}. It remains to evaluate the last integral. We do this by once more making the substitution $w_{nk}=z(k)\sqrt{g(k)}$ where $z(k)=e^{i\varphi(k)}/\sqrt{I_{nn}(k)}$. The integral is then simply given by

\begin{equation}
\begin{split}
    \int_{k} w_{nk}^\ast \partial_i\partial_j w_{nk}I_{nn}(k)
    =& \int_k \frac{\sqrt{g(k)}}{z(k)}\partial_i \partial_j (\sqrt{g(k)}z(k))\\
    =& \int_k g(k)\frac{\partial_i\partial_j z(k)}{z(k)}
    + \sqrt{g(k)}\partial_i\sqrt{g(k)}\frac{\partial_j z(k)}{z(k)}
    + \sqrt{g(k)}\partial_j\sqrt{g(k)}\frac{\partial_i z(k)}{z(k)}
    + \sqrt{g(k)}\partial_i \partial_j\sqrt{g(k)}\\
    =& \int_k g(k)\left( \frac{\partial_i\partial_j z(k)}{z(k)}
    -\frac{1}{2}\partial_i \frac{\partial_j z(k)}{z(k)}
    -\frac{1}{2}\partial_j \frac{\partial_i z(k)}{z(k)} \right)
    -\int_k \partial_i\sqrt{g(k)} \partial_j\sqrt{g(k)}\\
    =& - \alpha_i(k_c)\alpha_j(k_c) -\int_k \partial_i\sqrt{g(k)} \partial_j\sqrt{g(k)}.
\end{split}
\end{equation}
\end{widetext}
Plugging this result into \eqref{eq:rPr}, combining with \eqref{eq:rQr_final}, and making use of \eqref{eq:QGT_split}, \eqref{eq:Pnr}, \eqref{eq:r} we finally arrive at
\begin{equation}
    \Sigma_{ij} = g_{ij}(k_c) + \int_k \partial_i\sqrt{g(k)} \partial_j\sqrt{g(k)}.
\end{equation}

We see here that the covariance, which quantifies the real-space localization of the wavepacket, is indeed the sum of the quantum metric at the wavepacket center together with another matrix that depends only on the shape of the envelope function
\begin{equation}\label{eq:SigmaBar}
    \overline{\Sigma}_{ij}=\int_k \partial_i\sqrt{g(k)} \partial_j\sqrt{g(k)}.
\end{equation}
In this form, it is easy to see that, similarly to $\Sigma$ and $g$, $\overline{\Sigma}$ is also symmetric and, although less obvious, positive-definite. Positive-definiteness can be seen by considering that, for any vector $v$, we have
\begin{equation}
    v^T\overline{\Sigma}v=\braket{(v\cdot\nabla_k)\sqrt{g} \vert (v\cdot\nabla_k)\sqrt{g}}\geq0,
\end{equation}
and equality only occurs when $v=0$. Indeed, for any $v\neq0$, $\sqrt{g}$ is in the kernel of $v\cdot\nabla_k$ if and only if it is constant along the lines defined by $v$, which contradicts condition \eqref{eq:peakCondition}. We therefore conclude that the quantum metric $g_{ij}(k_c)$ is indeed a lower bound on the real-space covariance of the wavepacket.

We emphasize that the entirety of the calculation up until this point has been general and independent of the particular details of the wavepacket. As can be seen from its explicit form, however, the actual value of the matrix $\overline{\Sigma}$ does depend on these details, and cannot be calculated for a general wavepacket. 

Nevertheless, if the exact form of $g(k)$ is known or assumed, the values of $\overline{\Sigma}$ may be obtained directly. As an example, we consider the specific case of a Gaussian packet. That is, we assume that the envelope function has the following functional form:
\begin{equation}
    \label{eq:gaussian_envelope}
    g(k)=\frac{\exp\left({-\frac{1}{2}(k-k_c)^T\sigma^{-1}(k-k_c)}\right)}{\sqrt{(2\pi)^d\det\sigma}},
\end{equation}
where $d$ is the system dimension and $\sigma$ is the \emph{$k$-space} covariance matrix of the Gaussian. This envelope function satisfies condition \eqref{eq:peakCondition} in the limit $\sigma\to0$. Plugging this particular function into \eqref{eq:SigmaBar} yields
\begin{equation}\label{eq:cov_final}
    \overline{\Sigma}_{ij} = \frac{\sigma^{-1}_{ij}}{4}.
\end{equation}
As expected, this expression diverges for packets infinitely sharp in $k$-space.

\section{Numerical results in complex field}\label{App:graphs}

The numerical results for the field induced norm growth and velocity \eqref{eq:field_induced} were displayed in figures \ref{fig:N}, \ref{fig:r} respectively, covering the cases of purely real and purely imaginary field (parameter sets (I)-(VI) in table \ref{table:modelParams}). For the sake of completeness, we include here the results for the field induced norm growth and velocity in the complex field case (parameter sets (VII)-(IX)), presented in figure \ref{fig:complexfield}. We note that these results are nearly identical to the purely imaginary field case (parameter sets (IV)-(VI), respectively).

\bibliography{ComplexFieldEOM}

\begin{thebibliography}{82}%
\makeatletter
\providecommand \@ifxundefined [1]{%
 \@ifx{#1\undefined}
}%
\providecommand \@ifnum [1]{%
 \ifnum #1\expandafter \@firstoftwo
 \else \expandafter \@secondoftwo
 \fi
}%
\providecommand \@ifx [1]{%
 \ifx #1\expandafter \@firstoftwo
 \else \expandafter \@secondoftwo
 \fi
}%
\providecommand \natexlab [1]{#1}%
\providecommand \enquote  [1]{``#1''}%
\providecommand \bibnamefont  [1]{#1}%
\providecommand \bibfnamefont [1]{#1}%
\providecommand \citenamefont [1]{#1}%
\providecommand \href@noop [0]{\@secondoftwo}%
\providecommand \href [0]{\begingroup \@sanitize@url \@href}%
\providecommand \@href[1]{\@@startlink{#1}\@@href}%
\providecommand \@@href[1]{\endgroup#1\@@endlink}%
\providecommand \@sanitize@url [0]{\catcode `\\12\catcode `\$12\catcode
  `\&12\catcode `\#12\catcode `\^12\catcode `\_12\catcode `\%12\relax}%
\providecommand \@@startlink[1]{}%
\providecommand \@@endlink[0]{}%
\providecommand \url  [0]{\begingroup\@sanitize@url \@url }%
\providecommand \@url [1]{\endgroup\@href {#1}{\urlprefix }}%
\providecommand \urlprefix  [0]{URL }%
\providecommand \Eprint [0]{\href }%
\providecommand \doibase [0]{https://doi.org/}%
\providecommand \selectlanguage [0]{\@gobble}%
\providecommand \bibinfo  [0]{\@secondoftwo}%
\providecommand \bibfield  [0]{\@secondoftwo}%
\providecommand \translation [1]{[#1]}%
\providecommand \BibitemOpen [0]{}%
\providecommand \bibitemStop [0]{}%
\providecommand \bibitemNoStop [0]{.\EOS\space}%
\providecommand \EOS [0]{\spacefactor3000\relax}%
\providecommand \BibitemShut  [1]{\csname bibitem#1\endcsname}%
\let\auto@bib@innerbib\@empty
\bibitem [{\citenamefont {Berry}(1984)}]{berryQuantalPhaseFactors1984}%
  \BibitemOpen
  \bibfield  {author} {\bibinfo {author} {\bibfnamefont {M.~V.}\ \bibnamefont
  {Berry}},\ }\bibfield  {title} {\bibinfo {title} {Quantal phase factors
  accompanying adiabatic changes},\ }\href
  {https://doi.org/10.1098/rspa.1984.0023} {\bibfield  {journal} {\bibinfo
  {journal} {Proceedings of the Royal Society of London. A. Mathematical and
  Physical Sciences}\ }\textbf {\bibinfo {volume} {392}},\ \bibinfo {pages}
  {45} (\bibinfo {year} {1984})}\BibitemShut {NoStop}%
\bibitem [{\citenamefont {Zak}(1989)}]{ZakPhysRevLett.62.2747}%
  \BibitemOpen
  \bibfield  {author} {\bibinfo {author} {\bibfnamefont {J.}~\bibnamefont
  {Zak}},\ }\bibfield  {title} {\bibinfo {title} {Berry's phase for energy
  bands in solids},\ }\href {https://doi.org/10.1103/PhysRevLett.62.2747}
  {\bibfield  {journal} {\bibinfo  {journal} {Phys. Rev. Lett.}\ }\textbf
  {\bibinfo {volume} {62}},\ \bibinfo {pages} {2747} (\bibinfo {year}
  {1989})}\BibitemShut {NoStop}%
\bibitem [{\citenamefont {Meyer}\ \emph {et~al.}(2009)\citenamefont {Meyer},
  \citenamefont {Leanhardt}, \citenamefont {Cornell},\ and\ \citenamefont
  {Bohn}}]{MeyerPhysRevA.80.062110}%
  \BibitemOpen
  \bibfield  {author} {\bibinfo {author} {\bibfnamefont {E.~R.}\ \bibnamefont
  {Meyer}}, \bibinfo {author} {\bibfnamefont {A.~E.}\ \bibnamefont
  {Leanhardt}}, \bibinfo {author} {\bibfnamefont {E.~A.}\ \bibnamefont
  {Cornell}},\ and\ \bibinfo {author} {\bibfnamefont {J.~L.}\ \bibnamefont
  {Bohn}},\ }\bibfield  {title} {\bibinfo {title} {Berry-like phases in
  structured atoms and molecules},\ }\href
  {https://doi.org/10.1103/PhysRevA.80.062110} {\bibfield  {journal} {\bibinfo
  {journal} {Phys. Rev. A}\ }\textbf {\bibinfo {volume} {80}},\ \bibinfo
  {pages} {062110} (\bibinfo {year} {2009})}\BibitemShut {NoStop}%
\bibitem [{\citenamefont {Mironova}\ \emph {et~al.}(2013)\citenamefont
  {Mironova}, \citenamefont {Efremov},\ and\ \citenamefont
  {Schleich}}]{MironovaPhysRevA.87.013627}%
  \BibitemOpen
  \bibfield  {author} {\bibinfo {author} {\bibfnamefont {P.~V.}\ \bibnamefont
  {Mironova}}, \bibinfo {author} {\bibfnamefont {M.~A.}\ \bibnamefont
  {Efremov}},\ and\ \bibinfo {author} {\bibfnamefont {W.~P.}\ \bibnamefont
  {Schleich}},\ }\bibfield  {title} {\bibinfo {title} {Berry phase in atom
  optics},\ }\href {https://doi.org/10.1103/PhysRevA.87.013627} {\bibfield
  {journal} {\bibinfo  {journal} {Phys. Rev. A}\ }\textbf {\bibinfo {volume}
  {87}},\ \bibinfo {pages} {013627} (\bibinfo {year} {2013})}\BibitemShut
  {NoStop}%
\bibitem [{\citenamefont {Chiao}(1990)}]{chiaoBerryPhasesOptics1990}%
  \BibitemOpen
  \bibfield  {author} {\bibinfo {author} {\bibfnamefont {R.~Y.}\ \bibnamefont
  {Chiao}},\ }\bibfield  {title} {\bibinfo {title} {Berry's {{Phases}} in
  {{Optics}}},\ }in\ \href {https://doi.org/10.1007/978-94-009-2009-5_10}
  {\emph {\bibinfo {booktitle} {Analogies in {{Optics}} and {{Micro
  Electronics}}: {{Selected Contributions}} on {{Recent Developments}}}}},\
  \bibinfo {editor} {edited by\ \bibinfo {editor} {\bibfnamefont
  {W.}~\bibnamefont {{van Haeringen}}}\ and\ \bibinfo {editor} {\bibfnamefont
  {D.}~\bibnamefont {Lenstra}}}\ (\bibinfo  {publisher} {{Springer
  Netherlands}},\ \bibinfo {address} {{Dordrecht}},\ \bibinfo {year} {1990})\
  pp.\ \bibinfo {pages} {151--161}\BibitemShut {NoStop}%
\bibitem [{\citenamefont {Jisha}\ \emph {et~al.}(2021)\citenamefont {Jisha},
  \citenamefont {Nolte},\ and\ \citenamefont
  {Alberucci}}]{jishaGeometricPhaseOptics2021}%
  \BibitemOpen
  \bibfield  {author} {\bibinfo {author} {\bibfnamefont {C.~P.}\ \bibnamefont
  {Jisha}}, \bibinfo {author} {\bibfnamefont {S.}~\bibnamefont {Nolte}},\ and\
  \bibinfo {author} {\bibfnamefont {A.}~\bibnamefont {Alberucci}},\ }\bibfield
  {title} {\bibinfo {title} {Geometric {{Phase}} in {{Optics}}: {{From
  Wavefront Manipulation}} to {{Waveguiding}}},\ }\href
  {https://doi.org/10.1002/lpor.202100003} {\bibfield  {journal} {\bibinfo
  {journal} {Laser \& Photonics Reviews}\ }\textbf {\bibinfo {volume} {15}},\
  \bibinfo {pages} {2100003} (\bibinfo {year} {2021})}\BibitemShut {NoStop}%
\bibitem [{\citenamefont {Stone}\ \emph {et~al.}(2015)\citenamefont {Stone},
  \citenamefont {Dwivedi},\ and\ \citenamefont
  {Zhou}}]{StonePhysRevD.91.025004}%
  \BibitemOpen
  \bibfield  {author} {\bibinfo {author} {\bibfnamefont {M.}~\bibnamefont
  {Stone}}, \bibinfo {author} {\bibfnamefont {V.}~\bibnamefont {Dwivedi}},\
  and\ \bibinfo {author} {\bibfnamefont {T.}~\bibnamefont {Zhou}},\ }\bibfield
  {title} {\bibinfo {title} {{Berry} phase, {Lorentz} covariance, and anomalous
  velocity for {Dirac} and {Weyl} particles},\ }\href
  {https://doi.org/10.1103/PhysRevD.91.025004} {\bibfield  {journal} {\bibinfo
  {journal} {Phys. Rev. D}\ }\textbf {\bibinfo {volume} {91}},\ \bibinfo
  {pages} {025004} (\bibinfo {year} {2015})}\BibitemShut {NoStop}%
\bibitem [{\citenamefont {Baggio}\ \emph {et~al.}(2017)\citenamefont {Baggio},
  \citenamefont {Niarchos},\ and\ \citenamefont
  {Papadodimas}}]{baggioAspectsBerryPhase2017}%
  \BibitemOpen
  \bibfield  {author} {\bibinfo {author} {\bibfnamefont {M.}~\bibnamefont
  {Baggio}}, \bibinfo {author} {\bibfnamefont {V.}~\bibnamefont {Niarchos}},\
  and\ \bibinfo {author} {\bibfnamefont {K.}~\bibnamefont {Papadodimas}},\
  }\bibfield  {title} {\bibinfo {title} {Aspects of {{Berry}} phase in
  {{QFT}}},\ }\href {https://doi.org/10.1007/JHEP04(2017)062} {\bibfield
  {journal} {\bibinfo  {journal} {J. High Energ. Phys.}\ }\textbf {\bibinfo
  {volume} {2017}}\bibinfo  {number} { (4)},\ \bibinfo {pages}
  {62}}\BibitemShut {NoStop}%
\bibitem [{\citenamefont {King-Smith}\ and\ \citenamefont
  {Vanderbilt}(1993)}]{VanderbiltPhysRevB.47.1651}%
  \BibitemOpen
\bibfield  {number} {  }\bibfield  {author} {\bibinfo {author} {\bibfnamefont
  {R.~D.}\ \bibnamefont {King-Smith}}\ and\ \bibinfo {author} {\bibfnamefont
  {D.}~\bibnamefont {Vanderbilt}},\ }\bibfield  {title} {\bibinfo {title}
  {Theory of polarization of crystalline solids},\ }\href
  {https://doi.org/10.1103/PhysRevB.47.1651} {\bibfield  {journal} {\bibinfo
  {journal} {Phys. Rev. B}\ }\textbf {\bibinfo {volume} {47}},\ \bibinfo
  {pages} {1651} (\bibinfo {year} {1993})}\BibitemShut {NoStop}%
\bibitem [{\citenamefont {Ortiz}\ and\ \citenamefont
  {Martin}(1994)}]{OrtizPhysRevB.49.14202}%
  \BibitemOpen
  \bibfield  {author} {\bibinfo {author} {\bibfnamefont {G.}~\bibnamefont
  {Ortiz}}\ and\ \bibinfo {author} {\bibfnamefont {R.~M.}\ \bibnamefont
  {Martin}},\ }\bibfield  {title} {\bibinfo {title} {Macroscopic polarization
  as a geometric quantum phase: Many-body formulation},\ }\href
  {https://doi.org/10.1103/PhysRevB.49.14202} {\bibfield  {journal} {\bibinfo
  {journal} {Phys. Rev. B}\ }\textbf {\bibinfo {volume} {49}},\ \bibinfo
  {pages} {14202} (\bibinfo {year} {1994})}\BibitemShut {NoStop}%
\bibitem [{\citenamefont {Resta}(1994)}]{RestaRevModPhys.66.899}%
  \BibitemOpen
  \bibfield  {author} {\bibinfo {author} {\bibfnamefont {R.}~\bibnamefont
  {Resta}},\ }\bibfield  {title} {\bibinfo {title} {Macroscopic polarization in
  crystalline dielectrics: the geometric phase approach},\ }\href
  {https://doi.org/10.1103/RevModPhys.66.899} {\bibfield  {journal} {\bibinfo
  {journal} {Rev. Mod. Phys.}\ }\textbf {\bibinfo {volume} {66}},\ \bibinfo
  {pages} {899} (\bibinfo {year} {1994})}\BibitemShut {NoStop}%
\bibitem [{\citenamefont {Chang}\ and\ \citenamefont
  {Niu}(1996)}]{chang_berry_1996}%
  \BibitemOpen
  \bibfield  {author} {\bibinfo {author} {\bibfnamefont {M.-C.}\ \bibnamefont
  {Chang}}\ and\ \bibinfo {author} {\bibfnamefont {Q.}~\bibnamefont {Niu}},\
  }\bibfield  {title} {{\selectlanguage {english}\bibinfo {title} {Berry phase,
  hyperorbits, and the {Hofstadter} spectrum: {Semiclassical} dynamics in
  magnetic {Bloch} bands}},\ }\href {https://doi.org/10.1103/PhysRevB.53.7010}
  {\bibfield  {journal} {\bibinfo  {journal} {Physical Review B}\ }\textbf
  {\bibinfo {volume} {53}},\ \bibinfo {pages} {7010} (\bibinfo {year}
  {1996})}\BibitemShut {NoStop}%
\bibitem [{\citenamefont {Thonhauser}\ \emph {et~al.}(2005)\citenamefont
  {Thonhauser}, \citenamefont {Ceresoli}, \citenamefont {Vanderbilt},\ and\
  \citenamefont {Resta}}]{ThonhauserPhysRevLett.95.137205}%
  \BibitemOpen
  \bibfield  {author} {\bibinfo {author} {\bibfnamefont {T.}~\bibnamefont
  {Thonhauser}}, \bibinfo {author} {\bibfnamefont {D.}~\bibnamefont
  {Ceresoli}}, \bibinfo {author} {\bibfnamefont {D.}~\bibnamefont
  {Vanderbilt}},\ and\ \bibinfo {author} {\bibfnamefont {R.}~\bibnamefont
  {Resta}},\ }\bibfield  {title} {\bibinfo {title} {Orbital magnetization in
  periodic insulators},\ }\href {https://doi.org/10.1103/PhysRevLett.95.137205}
  {\bibfield  {journal} {\bibinfo  {journal} {Phys. Rev. Lett.}\ }\textbf
  {\bibinfo {volume} {95}},\ \bibinfo {pages} {137205} (\bibinfo {year}
  {2005})}\BibitemShut {NoStop}%
\bibitem [{\citenamefont {Thouless}\ \emph {et~al.}(1982)\citenamefont
  {Thouless}, \citenamefont {Kohmoto}, \citenamefont {Nightingale},\ and\
  \citenamefont {den Nijs}}]{ThoulessPhysRevLett.49.405}%
  \BibitemOpen
  \bibfield  {author} {\bibinfo {author} {\bibfnamefont {D.~J.}\ \bibnamefont
  {Thouless}}, \bibinfo {author} {\bibfnamefont {M.}~\bibnamefont {Kohmoto}},
  \bibinfo {author} {\bibfnamefont {M.~P.}\ \bibnamefont {Nightingale}},\ and\
  \bibinfo {author} {\bibfnamefont {M.}~\bibnamefont {den Nijs}},\ }\bibfield
  {title} {\bibinfo {title} {Quantized {Hall} conductance in a two-dimensional
  periodic potential},\ }\href {https://doi.org/10.1103/PhysRevLett.49.405}
  {\bibfield  {journal} {\bibinfo  {journal} {Phys. Rev. Lett.}\ }\textbf
  {\bibinfo {volume} {49}},\ \bibinfo {pages} {405} (\bibinfo {year}
  {1982})}\BibitemShut {NoStop}%
\bibitem [{\citenamefont {Haldane}(1988)}]{HaldanePhysRevLett.61.2015}%
  \BibitemOpen
  \bibfield  {author} {\bibinfo {author} {\bibfnamefont {F.~D.~M.}\
  \bibnamefont {Haldane}},\ }\bibfield  {title} {\bibinfo {title} {Model for a
  {Quantum Hall Effect} without {Landau Levels}: Condensed-matter realization
  of the "parity anomaly"},\ }\href
  {https://doi.org/10.1103/PhysRevLett.61.2015} {\bibfield  {journal} {\bibinfo
   {journal} {Phys. Rev. Lett.}\ }\textbf {\bibinfo {volume} {61}},\ \bibinfo
  {pages} {2015} (\bibinfo {year} {1988})}\BibitemShut {NoStop}%
\bibitem [{\citenamefont {Chang}\ and\ \citenamefont
  {Niu}(1995)}]{chang_berry_1995}%
  \BibitemOpen
  \bibfield  {author} {\bibinfo {author} {\bibfnamefont {M.-C.}\ \bibnamefont
  {Chang}}\ and\ \bibinfo {author} {\bibfnamefont {Q.}~\bibnamefont {Niu}},\
  }\bibfield  {title} {{\selectlanguage {english}\bibinfo {title} {Berry
  {Phase}, {Hyperorbits}, and the {Hofstadter} {Spectrum}}},\ }\href
  {https://doi.org/10.1103/PhysRevLett.75.1348} {\bibfield  {journal} {\bibinfo
   {journal} {Physical Review Letters}\ }\textbf {\bibinfo {volume} {75}},\
  \bibinfo {pages} {1348} (\bibinfo {year} {1995})}\BibitemShut {NoStop}%
\bibitem [{\citenamefont {Sundaram}\ and\ \citenamefont
  {Niu}(1999)}]{sundaram_wave-packet_1999}%
  \BibitemOpen
  \bibfield  {author} {\bibinfo {author} {\bibfnamefont {G.}~\bibnamefont
  {Sundaram}}\ and\ \bibinfo {author} {\bibfnamefont {Q.}~\bibnamefont {Niu}},\
  }\bibfield  {title} {{\selectlanguage {english}\bibinfo {title} {Wave-packet
  dynamics in slowly perturbed crystals: {Gradient} corrections and
  {Berry}-phase effects}},\ }\href {https://doi.org/10.1103/PhysRevB.59.14915}
  {\bibfield  {journal} {\bibinfo  {journal} {Physical Review B}\ }\textbf
  {\bibinfo {volume} {59}},\ \bibinfo {pages} {14915} (\bibinfo {year}
  {1999})}\BibitemShut {NoStop}%
\bibitem [{\citenamefont {Chang}\ and\ \citenamefont
  {Niu}(2008)}]{chang_berry_2008}%
  \BibitemOpen
  \bibfield  {author} {\bibinfo {author} {\bibfnamefont {M.-C.}\ \bibnamefont
  {Chang}}\ and\ \bibinfo {author} {\bibfnamefont {Q.}~\bibnamefont {Niu}},\
  }\bibfield  {title} {{\selectlanguage {english}\bibinfo {title} {Berry
  curvature, orbital moment, and effective quantum theory of electrons in
  electromagnetic fields}},\ }\href
  {https://doi.org/10.1088/0953-8984/20/19/193202} {\bibfield  {journal}
  {\bibinfo  {journal} {Journal of Physics: Condensed Matter}\ }\textbf
  {\bibinfo {volume} {20}},\ \bibinfo {pages} {193202} (\bibinfo {year}
  {2008})}\BibitemShut {NoStop}%
\bibitem [{\citenamefont {Xiao}\ \emph {et~al.}(2010)\citenamefont {Xiao},
  \citenamefont {Chang},\ and\ \citenamefont {Niu}}]{xiao_berry_2010}%
  \BibitemOpen
  \bibfield  {author} {\bibinfo {author} {\bibfnamefont {D.}~\bibnamefont
  {Xiao}}, \bibinfo {author} {\bibfnamefont {M.-C.}\ \bibnamefont {Chang}},\
  and\ \bibinfo {author} {\bibfnamefont {Q.}~\bibnamefont {Niu}},\ }\bibfield
  {title} {{\selectlanguage {english}\bibinfo {title} {Berry phase effects on
  electronic properties}},\ }\href {https://doi.org/10.1103/RevModPhys.82.1959}
  {\bibfield  {journal} {\bibinfo  {journal} {Reviews of Modern Physics}\
  }\textbf {\bibinfo {volume} {82}},\ \bibinfo {pages} {1959} (\bibinfo {year}
  {2010})}\BibitemShut {NoStop}%
\bibitem [{\citenamefont {Jungwirth}\ \emph {et~al.}(2002)\citenamefont
  {Jungwirth}, \citenamefont {Niu},\ and\ \citenamefont
  {MacDonald}}]{PhysRevLett.88.207208}%
  \BibitemOpen
  \bibfield  {author} {\bibinfo {author} {\bibfnamefont {T.}~\bibnamefont
  {Jungwirth}}, \bibinfo {author} {\bibfnamefont {Q.}~\bibnamefont {Niu}},\
  and\ \bibinfo {author} {\bibfnamefont {A.~H.}\ \bibnamefont {MacDonald}},\
  }\bibfield  {title} {\bibinfo {title} {Anomalous {Hall} effect in
  ferromagnetic semiconductors},\ }\href
  {https://doi.org/10.1103/PhysRevLett.88.207208} {\bibfield  {journal}
  {\bibinfo  {journal} {Phys. Rev. Lett.}\ }\textbf {\bibinfo {volume} {88}},\
  \bibinfo {pages} {207208} (\bibinfo {year} {2002})}\BibitemShut {NoStop}%
\bibitem [{\citenamefont {Nagaosa}\ \emph {et~al.}(2010)\citenamefont
  {Nagaosa}, \citenamefont {Sinova}, \citenamefont {Onoda}, \citenamefont
  {MacDonald},\ and\ \citenamefont {Ong}}]{RevModPhys.82.1539}%
  \BibitemOpen
  \bibfield  {author} {\bibinfo {author} {\bibfnamefont {N.}~\bibnamefont
  {Nagaosa}}, \bibinfo {author} {\bibfnamefont {J.}~\bibnamefont {Sinova}},
  \bibinfo {author} {\bibfnamefont {S.}~\bibnamefont {Onoda}}, \bibinfo
  {author} {\bibfnamefont {A.~H.}\ \bibnamefont {MacDonald}},\ and\ \bibinfo
  {author} {\bibfnamefont {N.~P.}\ \bibnamefont {Ong}},\ }\bibfield  {title}
  {\bibinfo {title} {Anomalous {Hall} effect},\ }\href
  {https://doi.org/10.1103/RevModPhys.82.1539} {\bibfield  {journal} {\bibinfo
  {journal} {Rev. Mod. Phys.}\ }\textbf {\bibinfo {volume} {82}},\ \bibinfo
  {pages} {1539} (\bibinfo {year} {2010})}\BibitemShut {NoStop}%
\bibitem [{\citenamefont {Provost}\ and\ \citenamefont
  {Vallee}(1980)}]{provost_riemannian_1980}%
  \BibitemOpen
  \bibfield  {author} {\bibinfo {author} {\bibfnamefont {J.~P.}\ \bibnamefont
  {Provost}}\ and\ \bibinfo {author} {\bibfnamefont {G.}~\bibnamefont
  {Vallee}},\ }\bibfield  {title} {{\selectlanguage {english}\bibinfo {title}
  {Riemannian structure on manifolds of quantum states}},\ }\href
  {https://doi.org/10.1007/BF02193559} {\bibfield  {journal} {\bibinfo
  {journal} {Communications in Mathematical Physics}\ }\textbf {\bibinfo
  {volume} {76}},\ \bibinfo {pages} {289} (\bibinfo {year} {1980})}\BibitemShut
  {NoStop}%
\bibitem [{\citenamefont {Resta}(2011)}]{resta_insulating_2011}%
  \BibitemOpen
  \bibfield  {author} {\bibinfo {author} {\bibfnamefont {R.}~\bibnamefont
  {Resta}},\ }\bibfield  {title} {{\selectlanguage {english}\bibinfo {title}
  {The insulating state of matter: a geometrical theory}},\ }\href
  {https://doi.org/10.1140/epjb/e2010-10874-4} {\bibfield  {journal} {\bibinfo
  {journal} {The European Physical Journal B}\ }\textbf {\bibinfo {volume}
  {79}},\ \bibinfo {pages} {121} (\bibinfo {year} {2011})}\BibitemShut
  {NoStop}%
\bibitem [{\citenamefont {Cayssol}\ and\ \citenamefont
  {Fuchs}(2021)}]{cayssol_topological_2021}%
  \BibitemOpen
  \bibfield  {author} {\bibinfo {author} {\bibfnamefont {J.}~\bibnamefont
  {Cayssol}}\ and\ \bibinfo {author} {\bibfnamefont {J.~N.}\ \bibnamefont
  {Fuchs}},\ }\bibfield  {title} {{\selectlanguage {english}\bibinfo {title}
  {Topological and geometrical aspects of band theory}},\ }\href
  {https://doi.org/10.1088/2515-7639/abf0b5} {\bibfield  {journal} {\bibinfo
  {journal} {Journal of Physics: Materials}\ }\textbf {\bibinfo {volume} {4}},\
  \bibinfo {pages} {034007} (\bibinfo {year} {2021})}\BibitemShut {NoStop}%
\bibitem [{\citenamefont {Liang}\ \emph {et~al.}(2017)\citenamefont {Liang},
  \citenamefont {Peotta}, \citenamefont {Harju},\ and\ \citenamefont
  {T\"orm\"a}}]{LiangPhysRevB.96.064511}%
  \BibitemOpen
  \bibfield  {author} {\bibinfo {author} {\bibfnamefont {L.}~\bibnamefont
  {Liang}}, \bibinfo {author} {\bibfnamefont {S.}~\bibnamefont {Peotta}},
  \bibinfo {author} {\bibfnamefont {A.}~\bibnamefont {Harju}},\ and\ \bibinfo
  {author} {\bibfnamefont {P.}~\bibnamefont {T\"orm\"a}},\ }\bibfield  {title}
  {\bibinfo {title} {Wave-packet dynamics of {Bogoliubov} quasiparticles:
  {Quantum} metric effects},\ }\href
  {https://doi.org/10.1103/PhysRevB.96.064511} {\bibfield  {journal} {\bibinfo
  {journal} {Phys. Rev. B}\ }\textbf {\bibinfo {volume} {96}},\ \bibinfo
  {pages} {064511} (\bibinfo {year} {2017})}\BibitemShut {NoStop}%
\bibitem [{\citenamefont {Gao}\ \emph {et~al.}(2014)\citenamefont {Gao},
  \citenamefont {Yang},\ and\ \citenamefont
  {Niu}}]{Gao_PhysRevLett.112.166601}%
  \BibitemOpen
  \bibfield  {author} {\bibinfo {author} {\bibfnamefont {Y.}~\bibnamefont
  {Gao}}, \bibinfo {author} {\bibfnamefont {S.~A.}\ \bibnamefont {Yang}},\ and\
  \bibinfo {author} {\bibfnamefont {Q.}~\bibnamefont {Niu}},\ }\bibfield
  {title} {\bibinfo {title} {Field induced positional shift of {Bloch}
  electrons and its dynamical implications},\ }\href
  {https://doi.org/10.1103/PhysRevLett.112.166601} {\bibfield  {journal}
  {\bibinfo  {journal} {Phys. Rev. Lett.}\ }\textbf {\bibinfo {volume} {112}},\
  \bibinfo {pages} {166601} (\bibinfo {year} {2014})}\BibitemShut {NoStop}%
\bibitem [{\citenamefont {Bleu}\ \emph {et~al.}(2018)\citenamefont {Bleu},
  \citenamefont {Malpuech}, \citenamefont {Gao},\ and\ \citenamefont
  {Solnyshkov}}]{bleu_effective_2018}%
  \BibitemOpen
  \bibfield  {author} {\bibinfo {author} {\bibfnamefont {O.}~\bibnamefont
  {Bleu}}, \bibinfo {author} {\bibfnamefont {G.}~\bibnamefont {Malpuech}},
  \bibinfo {author} {\bibfnamefont {Y.}~\bibnamefont {Gao}},\ and\ \bibinfo
  {author} {\bibfnamefont {D.~D.}\ \bibnamefont {Solnyshkov}},\ }\bibfield
  {title} {{\selectlanguage {english}\bibinfo {title} {Effective {Theory} of
  {Nonadiabatic} {Quantum} {Evolution} {Based} on the {Quantum} {Geometric}
  {Tensor}}},\ }\href {https://doi.org/10.1103/PhysRevLett.121.020401}
  {\bibfield  {journal} {\bibinfo  {journal} {Physical Review Letters}\
  }\textbf {\bibinfo {volume} {121}},\ \bibinfo {pages} {020401} (\bibinfo
  {year} {2018})}\BibitemShut {NoStop}%
\bibitem [{\citenamefont {Holder}\ \emph {et~al.}(2020)\citenamefont {Holder},
  \citenamefont {Kaplan},\ and\ \citenamefont
  {Yan}}]{Holder_PhysRevResearch.2.033100}%
  \BibitemOpen
  \bibfield  {author} {\bibinfo {author} {\bibfnamefont {T.}~\bibnamefont
  {Holder}}, \bibinfo {author} {\bibfnamefont {D.}~\bibnamefont {Kaplan}},\
  and\ \bibinfo {author} {\bibfnamefont {B.}~\bibnamefont {Yan}},\ }\bibfield
  {title} {\bibinfo {title} {Consequences of time-reversal-symmetry breaking in
  the light-matter interaction: Berry curvature, quantum metric, and diabatic
  motion},\ }\href {https://doi.org/10.1103/PhysRevResearch.2.033100}
  {\bibfield  {journal} {\bibinfo  {journal} {Phys. Rev. Res.}\ }\textbf
  {\bibinfo {volume} {2}},\ \bibinfo {pages} {033100} (\bibinfo {year}
  {2020})}\BibitemShut {NoStop}%
\bibitem [{\citenamefont {Leblanc}\ \emph {et~al.}(2021)\citenamefont
  {Leblanc}, \citenamefont {Malpuech},\ and\ \citenamefont
  {Solnyshkov}}]{leblanc_universal_2021}%
  \BibitemOpen
  \bibfield  {author} {\bibinfo {author} {\bibfnamefont {C.}~\bibnamefont
  {Leblanc}}, \bibinfo {author} {\bibfnamefont {G.}~\bibnamefont {Malpuech}},\
  and\ \bibinfo {author} {\bibfnamefont {D.~D.}\ \bibnamefont {Solnyshkov}},\
  }\bibfield  {title} {{\selectlanguage {english}\bibinfo {title} {Universal
  semiclassical equations based on the quantum metric for a two-band system}},\
  }\href {https://doi.org/10.1103/PhysRevB.104.134312} {\bibfield  {journal}
  {\bibinfo  {journal} {Physical Review B}\ }\textbf {\bibinfo {volume}
  {104}},\ \bibinfo {pages} {134312} (\bibinfo {year} {2021})}\BibitemShut
  {NoStop}%
\bibitem [{\citenamefont {Neupert}\ \emph {et~al.}(2013)\citenamefont
  {Neupert}, \citenamefont {Chamon},\ and\ \citenamefont
  {Mudry}}]{NeupertPhysRevB.87.245103}%
  \BibitemOpen
  \bibfield  {author} {\bibinfo {author} {\bibfnamefont {T.}~\bibnamefont
  {Neupert}}, \bibinfo {author} {\bibfnamefont {C.}~\bibnamefont {Chamon}},\
  and\ \bibinfo {author} {\bibfnamefont {C.}~\bibnamefont {Mudry}},\ }\bibfield
   {title} {\bibinfo {title} {Measuring the quantum geometry of {Bloch} bands
  with current noise},\ }\href {https://doi.org/10.1103/PhysRevB.87.245103}
  {\bibfield  {journal} {\bibinfo  {journal} {Phys. Rev. B}\ }\textbf {\bibinfo
  {volume} {87}},\ \bibinfo {pages} {245103} (\bibinfo {year}
  {2013})}\BibitemShut {NoStop}%
\bibitem [{\citenamefont {Ozawa}(2018)}]{ozawa_steady-state_2018}%
  \BibitemOpen
  \bibfield  {author} {\bibinfo {author} {\bibfnamefont {T.}~\bibnamefont
  {Ozawa}},\ }\bibfield  {title} {{\selectlanguage {english}\bibinfo {title}
  {Steady-state {Hall} response and quantum geometry of driven-dissipative
  lattices}},\ }\href {https://doi.org/10.1103/PhysRevB.97.041108} {\bibfield
  {journal} {\bibinfo  {journal} {Physical Review B}\ }\textbf {\bibinfo
  {volume} {97}},\ \bibinfo {pages} {041108(R)} (\bibinfo {year}
  {2018})}\BibitemShut {NoStop}%
\bibitem [{\citenamefont {Marzari}\ and\ \citenamefont
  {Vanderbilt}(1997)}]{marzari_maximally-localized_1997}%
  \BibitemOpen
  \bibfield  {author} {\bibinfo {author} {\bibfnamefont {N.}~\bibnamefont
  {Marzari}}\ and\ \bibinfo {author} {\bibfnamefont {D.}~\bibnamefont
  {Vanderbilt}},\ }\bibfield  {title} {{\selectlanguage {english}\bibinfo
  {title} {Maximally-localized generalized {Wannier} functions for composite
  energy bands}},\ }\href {https://doi.org/10.1103/PhysRevB.56.12847}
  {\bibfield  {journal} {\bibinfo  {journal} {Physical Review B}\ }\textbf
  {\bibinfo {volume} {56}},\ \bibinfo {pages} {12847} (\bibinfo {year}
  {1997})}\BibitemShut {NoStop}%
\bibitem [{\citenamefont {Chen}\ \emph {et~al.}(2014)\citenamefont {Chen},
  \citenamefont {Upadhyaya},\ and\ \citenamefont
  {Vitelli}}]{VitellichenNonlinearConductionSolitons2014}%
  \BibitemOpen
  \bibfield  {author} {\bibinfo {author} {\bibfnamefont {B.~G.-g.}\
  \bibnamefont {Chen}}, \bibinfo {author} {\bibfnamefont {N.}~\bibnamefont
  {Upadhyaya}},\ and\ \bibinfo {author} {\bibfnamefont {V.}~\bibnamefont
  {Vitelli}},\ }\bibfield  {title} {\bibinfo {title} {Nonlinear conduction via
  solitons in a topological mechanical insulator},\ }\href
  {https://doi.org/10.1073/pnas.1405969111} {\bibfield  {journal} {\bibinfo
  {journal} {Proceedings of the National Academy of Sciences}\ }\textbf
  {\bibinfo {volume} {111}},\ \bibinfo {pages} {13004} (\bibinfo {year}
  {2014})}\BibitemShut {NoStop}%
\bibitem [{\citenamefont {Süsstrunk}\ and\ \citenamefont
  {Huber}(2015)}]{Roman2015science.aab0239}%
  \BibitemOpen
  \bibfield  {author} {\bibinfo {author} {\bibfnamefont {R.}~\bibnamefont
  {Süsstrunk}}\ and\ \bibinfo {author} {\bibfnamefont {S.~D.}\ \bibnamefont
  {Huber}},\ }\bibfield  {title} {\bibinfo {title} {Observation of phononic
  helical edge states in a mechanical topological insulator},\ }\href
  {https://doi.org/10.1126/science.aab0239} {\bibfield  {journal} {\bibinfo
  {journal} {Science}\ }\textbf {\bibinfo {volume} {349}},\ \bibinfo {pages}
  {47} (\bibinfo {year} {2015})}\BibitemShut {NoStop}%
\bibitem [{\citenamefont {Neder}\ \emph {et~al.}(2023)\citenamefont {Neder},
  \citenamefont {Sirote}, \citenamefont {Geva}, \citenamefont {Lahini},
  \citenamefont {Ilan},\ and\ \citenamefont {Shokef}}]{neder2023bloch}%
  \BibitemOpen
  \bibfield  {author} {\bibinfo {author} {\bibfnamefont {I.}~\bibnamefont
  {Neder}}, \bibinfo {author} {\bibfnamefont {C.}~\bibnamefont {Sirote}},
  \bibinfo {author} {\bibfnamefont {M.}~\bibnamefont {Geva}}, \bibinfo {author}
  {\bibfnamefont {Y.}~\bibnamefont {Lahini}}, \bibinfo {author} {\bibfnamefont
  {R.}~\bibnamefont {Ilan}},\ and\ \bibinfo {author} {\bibfnamefont
  {Y.}~\bibnamefont {Shokef}},\ }\href@noop {} {\bibinfo {title} {Bloch
  oscillations, {Landau-Zener} transition, and topological phase evolution in a
  pendula array}} (\bibinfo {year} {2023}),\ \Eprint
  {https://arxiv.org/abs/2305.19387} {arXiv:2305.19387 [cond-mat.mes-hall]}
  \BibitemShut {NoStop}%
\bibitem [{\citenamefont {Raghu}\ and\ \citenamefont
  {Haldane}(2008)}]{RaghuPhysRevA.78.033834}%
  \BibitemOpen
  \bibfield  {author} {\bibinfo {author} {\bibfnamefont {S.}~\bibnamefont
  {Raghu}}\ and\ \bibinfo {author} {\bibfnamefont {F.~D.~M.}\ \bibnamefont
  {Haldane}},\ }\bibfield  {title} {\bibinfo {title} {Analogs of
  quantum-hall-effect edge states in photonic crystals},\ }\href
  {https://doi.org/10.1103/PhysRevA.78.033834} {\bibfield  {journal} {\bibinfo
  {journal} {Phys. Rev. A}\ }\textbf {\bibinfo {volume} {78}},\ \bibinfo
  {pages} {033834} (\bibinfo {year} {2008})}\BibitemShut {NoStop}%
\bibitem [{\citenamefont {Wang}\ \emph {et~al.}(2009)\citenamefont {Wang},
  \citenamefont {Chong}, \citenamefont {Joannopoulos},\ and\ \citenamefont
  {Solja{\v
  c}i{\'c}}}]{wangObservationUnidirectionalBackscatteringimmune2009a}%
  \BibitemOpen
  \bibfield  {author} {\bibinfo {author} {\bibfnamefont {Z.}~\bibnamefont
  {Wang}}, \bibinfo {author} {\bibfnamefont {Y.}~\bibnamefont {Chong}},
  \bibinfo {author} {\bibfnamefont {J.~D.}\ \bibnamefont {Joannopoulos}},\ and\
  \bibinfo {author} {\bibfnamefont {M.}~\bibnamefont {Solja{\v c}i{\'c}}},\
  }\bibfield  {title} {\bibinfo {title} {Observation of unidirectional
  backscattering-immune topological electromagnetic states},\ }\href
  {https://doi.org/10.1038/nature08293} {\bibfield  {journal} {\bibinfo
  {journal} {Nature}\ }\textbf {\bibinfo {volume} {461}},\ \bibinfo {pages}
  {772} (\bibinfo {year} {2009})}\BibitemShut {NoStop}%
\bibitem [{\citenamefont {Ozawa}\ \emph {et~al.}(2019)\citenamefont {Ozawa},
  \citenamefont {Price}, \citenamefont {Amo}, \citenamefont {Goldman},
  \citenamefont {Hafezi}, \citenamefont {Lu}, \citenamefont {Rechtsman},
  \citenamefont {Schuster}, \citenamefont {Simon}, \citenamefont {Zilberberg},\
  and\ \citenamefont {Carusotto}}]{OzawaRevModPhys.91.015006}%
  \BibitemOpen
  \bibfield  {author} {\bibinfo {author} {\bibfnamefont {T.}~\bibnamefont
  {Ozawa}}, \bibinfo {author} {\bibfnamefont {H.~M.}\ \bibnamefont {Price}},
  \bibinfo {author} {\bibfnamefont {A.}~\bibnamefont {Amo}}, \bibinfo {author}
  {\bibfnamefont {N.}~\bibnamefont {Goldman}}, \bibinfo {author} {\bibfnamefont
  {M.}~\bibnamefont {Hafezi}}, \bibinfo {author} {\bibfnamefont
  {L.}~\bibnamefont {Lu}}, \bibinfo {author} {\bibfnamefont {M.~C.}\
  \bibnamefont {Rechtsman}}, \bibinfo {author} {\bibfnamefont {D.}~\bibnamefont
  {Schuster}}, \bibinfo {author} {\bibfnamefont {J.}~\bibnamefont {Simon}},
  \bibinfo {author} {\bibfnamefont {O.}~\bibnamefont {Zilberberg}},\ and\
  \bibinfo {author} {\bibfnamefont {I.}~\bibnamefont {Carusotto}},\ }\bibfield
  {title} {\bibinfo {title} {Topological photonics},\ }\href
  {https://doi.org/10.1103/RevModPhys.91.015006} {\bibfield  {journal}
  {\bibinfo  {journal} {Rev. Mod. Phys.}\ }\textbf {\bibinfo {volume} {91}},\
  \bibinfo {pages} {015006} (\bibinfo {year} {2019})}\BibitemShut {NoStop}%
\bibitem [{\citenamefont {He}\ \emph {et~al.}(2016)\citenamefont {He},
  \citenamefont {Ni}, \citenamefont {Ge}, \citenamefont {Sun}, \citenamefont
  {Chen}, \citenamefont {Lu}, \citenamefont {Liu},\ and\ \citenamefont
  {Chen}}]{heAcousticTopologicalInsulator2016}%
  \BibitemOpen
  \bibfield  {author} {\bibinfo {author} {\bibfnamefont {C.}~\bibnamefont
  {He}}, \bibinfo {author} {\bibfnamefont {X.}~\bibnamefont {Ni}}, \bibinfo
  {author} {\bibfnamefont {H.}~\bibnamefont {Ge}}, \bibinfo {author}
  {\bibfnamefont {X.-C.}\ \bibnamefont {Sun}}, \bibinfo {author} {\bibfnamefont
  {Y.-B.}\ \bibnamefont {Chen}}, \bibinfo {author} {\bibfnamefont {M.-H.}\
  \bibnamefont {Lu}}, \bibinfo {author} {\bibfnamefont {X.-P.}\ \bibnamefont
  {Liu}},\ and\ \bibinfo {author} {\bibfnamefont {Y.-F.}\ \bibnamefont
  {Chen}},\ }\bibfield  {title} {\bibinfo {title} {Acoustic topological
  insulator and robust one-way sound transport},\ }\href
  {https://doi.org/10.1038/nphys3867} {\bibfield  {journal} {\bibinfo
  {journal} {Nature Phys}\ }\textbf {\bibinfo {volume} {12}},\ \bibinfo {pages}
  {1124} (\bibinfo {year} {2016})}\BibitemShut {NoStop}%
\bibitem [{\citenamefont {Qi}\ \emph {et~al.}(2020)\citenamefont {Qi},
  \citenamefont {Qiu}, \citenamefont {Xiao}, \citenamefont {He}, \citenamefont
  {Ke},\ and\ \citenamefont {Liu}}]{QiPhysRevLett.124.206601}%
  \BibitemOpen
  \bibfield  {author} {\bibinfo {author} {\bibfnamefont {Y.}~\bibnamefont
  {Qi}}, \bibinfo {author} {\bibfnamefont {C.}~\bibnamefont {Qiu}}, \bibinfo
  {author} {\bibfnamefont {M.}~\bibnamefont {Xiao}}, \bibinfo {author}
  {\bibfnamefont {H.}~\bibnamefont {He}}, \bibinfo {author} {\bibfnamefont
  {M.}~\bibnamefont {Ke}},\ and\ \bibinfo {author} {\bibfnamefont
  {Z.}~\bibnamefont {Liu}},\ }\bibfield  {title} {\bibinfo {title} {Acoustic
  realization of quadrupole topological insulators},\ }\href
  {https://doi.org/10.1103/PhysRevLett.124.206601} {\bibfield  {journal}
  {\bibinfo  {journal} {Phys. Rev. Lett.}\ }\textbf {\bibinfo {volume} {124}},\
  \bibinfo {pages} {206601} (\bibinfo {year} {2020})}\BibitemShut {NoStop}%
\bibitem [{\citenamefont {Bender}(2007)}]{Bender_2007}%
  \BibitemOpen
  \bibfield  {author} {\bibinfo {author} {\bibfnamefont {C.~M.}\ \bibnamefont
  {Bender}},\ }\bibfield  {title} {\bibinfo {title} {Making sense of
  {non-Hermitian} {Hamiltonians}},\ }\href
  {https://doi.org/10.1088/0034-4885/70/6/R03} {\bibfield  {journal} {\bibinfo
  {journal} {Reports on Progress in Physics}\ }\textbf {\bibinfo {volume}
  {70}},\ \bibinfo {pages} {947} (\bibinfo {year} {2007})}\BibitemShut
  {NoStop}%
\bibitem [{\citenamefont {Shen}\ \emph {et~al.}(2018)\citenamefont {Shen},
  \citenamefont {Zhen},\ and\ \citenamefont {Fu}}]{ShenPhysRevLett.120.146402}%
  \BibitemOpen
  \bibfield  {author} {\bibinfo {author} {\bibfnamefont {H.}~\bibnamefont
  {Shen}}, \bibinfo {author} {\bibfnamefont {B.}~\bibnamefont {Zhen}},\ and\
  \bibinfo {author} {\bibfnamefont {L.}~\bibnamefont {Fu}},\ }\bibfield
  {title} {\bibinfo {title} {Topological band theory for non-hermitian
  hamiltonians},\ }\href {https://doi.org/10.1103/PhysRevLett.120.146402}
  {\bibfield  {journal} {\bibinfo  {journal} {Phys. Rev. Lett.}\ }\textbf
  {\bibinfo {volume} {120}},\ \bibinfo {pages} {146402} (\bibinfo {year}
  {2018})}\BibitemShut {NoStop}%
\bibitem [{\citenamefont {Martinez~Alvarez}\ \emph {et~al.}(2018)\citenamefont
  {Martinez~Alvarez}, \citenamefont {Barrios~Vargas}, \citenamefont
  {Berdakin},\ and\ \citenamefont
  {Foa~Torres}}]{martinezalvarezTopologicalStatesNonHermitian2018}%
  \BibitemOpen
  \bibfield  {author} {\bibinfo {author} {\bibfnamefont {V.~M.}\ \bibnamefont
  {Martinez~Alvarez}}, \bibinfo {author} {\bibfnamefont {J.~E.}\ \bibnamefont
  {Barrios~Vargas}}, \bibinfo {author} {\bibfnamefont {M.}~\bibnamefont
  {Berdakin}},\ and\ \bibinfo {author} {\bibfnamefont {L.~E.~F.}\ \bibnamefont
  {Foa~Torres}},\ }\bibfield  {title} {\bibinfo {title} {Topological states of
  non-{{Hermitian}} systems},\ }\href
  {https://doi.org/10.1140/epjst/e2018-800091-5} {\bibfield  {journal}
  {\bibinfo  {journal} {Eur. Phys. J. Spec. Top.}\ }\textbf {\bibinfo {volume}
  {227}},\ \bibinfo {pages} {1295} (\bibinfo {year} {2018})}\BibitemShut
  {NoStop}%
\bibitem [{\citenamefont {Ghatak}\ and\ \citenamefont
  {Das}(2019)}]{Ghatak_2019}%
  \BibitemOpen
  \bibfield  {author} {\bibinfo {author} {\bibfnamefont {A.}~\bibnamefont
  {Ghatak}}\ and\ \bibinfo {author} {\bibfnamefont {T.}~\bibnamefont {Das}},\
  }\bibfield  {title} {\bibinfo {title} {New topological invariants in
  non-hermitian systems},\ }\href {https://doi.org/10.1088/1361-648X/ab11b3}
  {\bibfield  {journal} {\bibinfo  {journal} {Journal of Physics: Condensed
  Matter}\ }\textbf {\bibinfo {volume} {31}},\ \bibinfo {pages} {263001}
  (\bibinfo {year} {2019})}\BibitemShut {NoStop}%
\bibitem [{\citenamefont {Kunst}\ and\ \citenamefont
  {Dwivedi}(2019)}]{KunstPhysRevB.99.245116}%
  \BibitemOpen
  \bibfield  {author} {\bibinfo {author} {\bibfnamefont {F.~K.}\ \bibnamefont
  {Kunst}}\ and\ \bibinfo {author} {\bibfnamefont {V.}~\bibnamefont
  {Dwivedi}},\ }\bibfield  {title} {\bibinfo {title} {Non-hermitian systems and
  topology: A transfer-matrix perspective},\ }\href
  {https://doi.org/10.1103/PhysRevB.99.245116} {\bibfield  {journal} {\bibinfo
  {journal} {Phys. Rev. B}\ }\textbf {\bibinfo {volume} {99}},\ \bibinfo
  {pages} {245116} (\bibinfo {year} {2019})}\BibitemShut {NoStop}%
\bibitem [{\citenamefont {Gong}\ \emph {et~al.}(2018)\citenamefont {Gong},
  \citenamefont {Ashida}, \citenamefont {Kawabata}, \citenamefont {Takasan},
  \citenamefont {Higashikawa},\ and\ \citenamefont
  {Ueda}}]{gong_topological_2018}%
  \BibitemOpen
  \bibfield  {author} {\bibinfo {author} {\bibfnamefont {Z.}~\bibnamefont
  {Gong}}, \bibinfo {author} {\bibfnamefont {Y.}~\bibnamefont {Ashida}},
  \bibinfo {author} {\bibfnamefont {K.}~\bibnamefont {Kawabata}}, \bibinfo
  {author} {\bibfnamefont {K.}~\bibnamefont {Takasan}}, \bibinfo {author}
  {\bibfnamefont {S.}~\bibnamefont {Higashikawa}},\ and\ \bibinfo {author}
  {\bibfnamefont {M.}~\bibnamefont {Ueda}},\ }\bibfield  {title} {\bibinfo
  {title} {Topological {Phases} of {Non}-{Hermitian} {Systems}},\ }\href
  {https://doi.org/10.1103/PhysRevX.8.031079} {\bibfield  {journal} {\bibinfo
  {journal} {Physical Review X}\ }\textbf {\bibinfo {volume} {8}},\ \bibinfo
  {pages} {031079} (\bibinfo {year} {2018})}\BibitemShut {NoStop}%
\bibitem [{\citenamefont {Ashida}\ \emph {et~al.}(2020)\citenamefont {Ashida},
  \citenamefont {Gong},\ and\ \citenamefont
  {Ueda}}]{ashida_non-hermitian_2020}%
  \BibitemOpen
  \bibfield  {author} {\bibinfo {author} {\bibfnamefont {Y.}~\bibnamefont
  {Ashida}}, \bibinfo {author} {\bibfnamefont {Z.}~\bibnamefont {Gong}},\ and\
  \bibinfo {author} {\bibfnamefont {M.}~\bibnamefont {Ueda}},\ }\bibfield
  {title} {{\selectlanguage {english}\bibinfo {title} {Non-{Hermitian}
  {Physics}}},\ }\href {https://doi.org/10.1080/00018732.2021.1876991}
  {\bibfield  {journal} {\bibinfo  {journal} {Advances in Physics}\ }\textbf
  {\bibinfo {volume} {69}},\ \bibinfo {pages} {249} (\bibinfo {year}
  {2020})}\BibitemShut {NoStop}%
\bibitem [{\citenamefont {Martínez-Strasser}\ \emph
  {et~al.}(2024)\citenamefont {Martínez-Strasser}, \citenamefont {Herrera},
  \citenamefont {García-Etxarri}, \citenamefont {Palumbo}, \citenamefont
  {Kunst},\ and\ \citenamefont
  {Bercioux}}]{Martinezhttps://doi.org/10.1002/qute.202300225}%
  \BibitemOpen
  \bibfield  {author} {\bibinfo {author} {\bibfnamefont {C.}~\bibnamefont
  {Martínez-Strasser}}, \bibinfo {author} {\bibfnamefont {M.~A.~J.}\
  \bibnamefont {Herrera}}, \bibinfo {author} {\bibfnamefont {A.}~\bibnamefont
  {García-Etxarri}}, \bibinfo {author} {\bibfnamefont {G.}~\bibnamefont
  {Palumbo}}, \bibinfo {author} {\bibfnamefont {F.~K.}\ \bibnamefont {Kunst}},\
  and\ \bibinfo {author} {\bibfnamefont {D.}~\bibnamefont {Bercioux}},\
  }\bibfield  {title} {\bibinfo {title} {Topological properties of a
  non-hermitian quasi-1d chain with a flat band},\ }\href
  {https://doi.org/https://doi.org/10.1002/qute.202300225} {\bibfield
  {journal} {\bibinfo  {journal} {Advanced Quantum Technologies}\ }\textbf
  {\bibinfo {volume} {7}},\ \bibinfo {pages} {2300225} (\bibinfo {year}
  {2024})}\BibitemShut {NoStop}%
\bibitem [{\citenamefont {Nenciu}\ and\ \citenamefont
  {Rasche}(1992)}]{GNenciu_1992}%
  \BibitemOpen
  \bibfield  {author} {\bibinfo {author} {\bibfnamefont {G.}~\bibnamefont
  {Nenciu}}\ and\ \bibinfo {author} {\bibfnamefont {G.}~\bibnamefont
  {Rasche}},\ }\bibfield  {title} {\bibinfo {title} {On the adiabatic theorem
  for nonself-adjoint {Hamiltonians}},\ }\href
  {https://doi.org/10.1088/0305-4470/25/21/027} {\bibfield  {journal} {\bibinfo
   {journal} {Journal of Physics A: Mathematical and General}\ }\textbf
  {\bibinfo {volume} {25}},\ \bibinfo {pages} {5741} (\bibinfo {year}
  {1992})}\BibitemShut {NoStop}%
\bibitem [{\citenamefont {Berry}\ and\ \citenamefont
  {Uzdin}(2011)}]{berry_slow_2011}%
  \BibitemOpen
  \bibfield  {author} {\bibinfo {author} {\bibfnamefont {M.~V.}\ \bibnamefont
  {Berry}}\ and\ \bibinfo {author} {\bibfnamefont {R.}~\bibnamefont {Uzdin}},\
  }\bibfield  {title} {{\selectlanguage {english}\bibinfo {title} {Slow
  non-{Hermitian} cycling: exact solutions and the {Stokes} phenomenon}},\
  }\href {https://doi.org/10.1088/1751-8113/44/43/435303} {\bibfield  {journal}
  {\bibinfo  {journal} {Journal of Physics A: Mathematical and Theoretical}\
  }\textbf {\bibinfo {volume} {44}},\ \bibinfo {pages} {435303} (\bibinfo
  {year} {2011})}\BibitemShut {NoStop}%
\bibitem [{\citenamefont {Ibáñez}\ and\ \citenamefont
  {Muga}(2014)}]{ibanez_adiabaticity_2014}%
  \BibitemOpen
  \bibfield  {author} {\bibinfo {author} {\bibfnamefont {S.}~\bibnamefont
  {Ibáñez}}\ and\ \bibinfo {author} {\bibfnamefont {J.~G.}\ \bibnamefont
  {Muga}},\ }\bibfield  {title} {\bibinfo {title} {Adiabaticity condition for
  non-{Hermitian} {Hamiltonians}},\ }\href
  {https://doi.org/10.1103/PhysRevA.89.033403} {\bibfield  {journal} {\bibinfo
  {journal} {Physical Review A}\ }\textbf {\bibinfo {volume} {89}},\ \bibinfo
  {pages} {033403} (\bibinfo {year} {2014})}\BibitemShut {NoStop}%
\bibitem [{\citenamefont {Wang}\ \emph {et~al.}(2018)\citenamefont {Wang},
  \citenamefont {Lang},\ and\ \citenamefont {Chong}}]{wang_non-hermitian_2018}%
  \BibitemOpen
  \bibfield  {author} {\bibinfo {author} {\bibfnamefont {H.}~\bibnamefont
  {Wang}}, \bibinfo {author} {\bibfnamefont {L.-J.}\ \bibnamefont {Lang}},\
  and\ \bibinfo {author} {\bibfnamefont {Y.~D.}\ \bibnamefont {Chong}},\
  }\bibfield  {title} {\bibinfo {title} {Non-{Hermitian} dynamics of slowly
  varying {Hamiltonians}},\ }\href {https://doi.org/10.1103/PhysRevA.98.012119}
  {\bibfield  {journal} {\bibinfo  {journal} {Physical Review A}\ }\textbf
  {\bibinfo {volume} {98}},\ \bibinfo {pages} {012119} (\bibinfo {year}
  {2018})}\BibitemShut {NoStop}%
\bibitem [{\citenamefont {Bender}\ and\ \citenamefont
  {Boettcher}(1998)}]{BenderPhysRevLett.80.5243}%
  \BibitemOpen
  \bibfield  {author} {\bibinfo {author} {\bibfnamefont {C.~M.}\ \bibnamefont
  {Bender}}\ and\ \bibinfo {author} {\bibfnamefont {S.}~\bibnamefont
  {Boettcher}},\ }\bibfield  {title} {\bibinfo {title} {Real spectra in
  {Non-Hermitian} {Hamiltonians} having $\mathcal{PT}$ symmetry},\ }\href
  {https://doi.org/10.1103/PhysRevLett.80.5243} {\bibfield  {journal} {\bibinfo
   {journal} {Phys. Rev. Lett.}\ }\textbf {\bibinfo {volume} {80}},\ \bibinfo
  {pages} {5243} (\bibinfo {year} {1998})}\BibitemShut {NoStop}%
\bibitem [{\citenamefont {Bender}\ and\ \citenamefont
  {Mannheim}(2010)}]{BENDER20101616}%
  \BibitemOpen
  \bibfield  {author} {\bibinfo {author} {\bibfnamefont {C.~M.}\ \bibnamefont
  {Bender}}\ and\ \bibinfo {author} {\bibfnamefont {P.~D.}\ \bibnamefont
  {Mannheim}},\ }\bibfield  {title} {\bibinfo {title} {{PT} symmetry and
  necessary and sufficient conditions for the reality of energy eigenvalues},\
  }\href {https://doi.org/https://doi.org/10.1016/j.physleta.2010.02.032}
  {\bibfield  {journal} {\bibinfo  {journal} {Physics Letters A}\ }\textbf
  {\bibinfo {volume} {374}},\ \bibinfo {pages} {1616} (\bibinfo {year}
  {2010})}\BibitemShut {NoStop}%
\bibitem [{\citenamefont {Brody}(2013)}]{Brody_2014}%
  \BibitemOpen
  \bibfield  {author} {\bibinfo {author} {\bibfnamefont {D.~C.}\ \bibnamefont
  {Brody}},\ }\bibfield  {title} {\bibinfo {title} {Biorthogonal quantum
  mechanics},\ }\href {https://doi.org/10.1088/1751-8113/47/3/035305}
  {\bibfield  {journal} {\bibinfo  {journal} {Journal of Physics A:
  Mathematical and Theoretical}\ }\textbf {\bibinfo {volume} {47}},\ \bibinfo
  {pages} {035305} (\bibinfo {year} {2013})}\BibitemShut {NoStop}%
\bibitem [{\citenamefont {Bergholtz}\ \emph {et~al.}(2021)\citenamefont
  {Bergholtz}, \citenamefont {Budich},\ and\ \citenamefont
  {Kunst}}]{bergholtz_exceptional_2021}%
  \BibitemOpen
  \bibfield  {author} {\bibinfo {author} {\bibfnamefont {E.~J.}\ \bibnamefont
  {Bergholtz}}, \bibinfo {author} {\bibfnamefont {J.~C.}\ \bibnamefont
  {Budich}},\ and\ \bibinfo {author} {\bibfnamefont {F.~K.}\ \bibnamefont
  {Kunst}},\ }\bibfield  {title} {\bibinfo {title} {Exceptional topology of
  non-{Hermitian} systems},\ }\href
  {https://doi.org/10.1103/RevModPhys.93.015005} {\bibfield  {journal}
  {\bibinfo  {journal} {Reviews of Modern Physics}\ }\textbf {\bibinfo {volume}
  {93}},\ \bibinfo {pages} {015005} (\bibinfo {year} {2021})}\BibitemShut
  {NoStop}%
\bibitem [{\citenamefont {Graefe}\ and\ \citenamefont
  {Schubert}(2011)}]{graefe_wave-packet_2011}%
  \BibitemOpen
  \bibfield  {author} {\bibinfo {author} {\bibfnamefont {E.-M.}\ \bibnamefont
  {Graefe}}\ and\ \bibinfo {author} {\bibfnamefont {R.}~\bibnamefont
  {Schubert}},\ }\bibfield  {title} {\bibinfo {title} {Wave-packet evolution in
  non-{Hermitian} quantum systems},\ }\href
  {https://doi.org/10.1103/PhysRevA.83.060101} {\bibfield  {journal} {\bibinfo
  {journal} {Physical Review A}\ }\textbf {\bibinfo {volume} {83}},\ \bibinfo
  {pages} {060101(R)} (\bibinfo {year} {2011})}\BibitemShut {NoStop}%
\bibitem [{\citenamefont {Schomerus}\ and\ \citenamefont
  {Wiersig}(2014)}]{schomerus_non-hermitian-transport_2014}%
  \BibitemOpen
  \bibfield  {author} {\bibinfo {author} {\bibfnamefont {H.}~\bibnamefont
  {Schomerus}}\ and\ \bibinfo {author} {\bibfnamefont {J.}~\bibnamefont
  {Wiersig}},\ }\bibfield  {title} {\bibinfo {title} {Non-{Hermitian}-transport
  effects in coupled-resonator optical waveguides},\ }\href
  {https://doi.org/10.1103/PhysRevA.90.053819} {\bibfield  {journal} {\bibinfo
  {journal} {Physical Review A}\ }\textbf {\bibinfo {volume} {90}},\ \bibinfo
  {pages} {053819} (\bibinfo {year} {2014})}\BibitemShut {NoStop}%
\bibitem [{\citenamefont {Xu}\ \emph {et~al.}(2017)\citenamefont {Xu},
  \citenamefont {Wang},\ and\ \citenamefont {Duan}}]{xu_weyl_2017}%
  \BibitemOpen
  \bibfield  {author} {\bibinfo {author} {\bibfnamefont {Y.}~\bibnamefont
  {Xu}}, \bibinfo {author} {\bibfnamefont {S.-T.}\ \bibnamefont {Wang}},\ and\
  \bibinfo {author} {\bibfnamefont {L.-M.}\ \bibnamefont {Duan}},\ }\bibfield
  {title} {\bibinfo {title} {Weyl {Exceptional} {Rings} in a
  {Three}-{Dimensional} {Dissipative} {Cold} {Atomic} {Gas}},\ }\href
  {https://doi.org/10.1103/PhysRevLett.118.045701} {\bibfield  {journal}
  {\bibinfo  {journal} {Physical Review Letters}\ }\textbf {\bibinfo {volume}
  {118}},\ \bibinfo {pages} {045701} (\bibinfo {year} {2017})}\BibitemShut
  {NoStop}%
\bibitem [{\citenamefont {Singhal}\ \emph {et~al.}(2023)\citenamefont
  {Singhal}, \citenamefont {Martello}, \citenamefont {Agrawal}, \citenamefont
  {Ozawa}, \citenamefont {Price},\ and\ \citenamefont
  {Gadway}}]{PhysRevResearch.5.L032026}%
  \BibitemOpen
  \bibfield  {author} {\bibinfo {author} {\bibfnamefont {Y.}~\bibnamefont
  {Singhal}}, \bibinfo {author} {\bibfnamefont {E.}~\bibnamefont {Martello}},
  \bibinfo {author} {\bibfnamefont {S.}~\bibnamefont {Agrawal}}, \bibinfo
  {author} {\bibfnamefont {T.}~\bibnamefont {Ozawa}}, \bibinfo {author}
  {\bibfnamefont {H.}~\bibnamefont {Price}},\ and\ \bibinfo {author}
  {\bibfnamefont {B.}~\bibnamefont {Gadway}},\ }\bibfield  {title} {\bibinfo
  {title} {Measuring the adiabatic {non-Hermitian Berry} phase in
  feedback-coupled oscillators},\ }\href
  {https://doi.org/10.1103/PhysRevResearch.5.L032026} {\bibfield  {journal}
  {\bibinfo  {journal} {Phys. Rev. Res.}\ }\textbf {\bibinfo {volume} {5}},\
  \bibinfo {pages} {L032026} (\bibinfo {year} {2023})}\BibitemShut {NoStop}%
\bibitem [{\citenamefont {Martello}\ \emph {et~al.}(2023)\citenamefont
  {Martello}, \citenamefont {Singhal}, \citenamefont {Gadway}, \citenamefont
  {Ozawa},\ and\ \citenamefont {Price}}]{martello_coexistence_2023}%
  \BibitemOpen
  \bibfield  {author} {\bibinfo {author} {\bibfnamefont {E.}~\bibnamefont
  {Martello}}, \bibinfo {author} {\bibfnamefont {Y.}~\bibnamefont {Singhal}},
  \bibinfo {author} {\bibfnamefont {B.}~\bibnamefont {Gadway}}, \bibinfo
  {author} {\bibfnamefont {T.}~\bibnamefont {Ozawa}},\ and\ \bibinfo {author}
  {\bibfnamefont {H.~M.}\ \bibnamefont {Price}},\ }\bibfield  {title} {\bibinfo
  {title} {Coexistence of stable and unstable population dynamics in a
  nonlinear non-{Hermitian} mechanical dimer},\ }\href
  {https://doi.org/10.1103/PhysRevE.107.064211} {\bibfield  {journal} {\bibinfo
   {journal} {Physical Review E}\ }\textbf {\bibinfo {volume} {107}},\ \bibinfo
  {pages} {064211} (\bibinfo {year} {2023})}\BibitemShut {NoStop}%
\bibitem [{\citenamefont {Silberstein}\ \emph {et~al.}(2020)\citenamefont
  {Silberstein}, \citenamefont {Behrends}, \citenamefont {Goldstein},\ and\
  \citenamefont {Ilan}}]{silberstein_berry_2020}%
  \BibitemOpen
  \bibfield  {author} {\bibinfo {author} {\bibfnamefont {N.}~\bibnamefont
  {Silberstein}}, \bibinfo {author} {\bibfnamefont {J.}~\bibnamefont
  {Behrends}}, \bibinfo {author} {\bibfnamefont {M.}~\bibnamefont
  {Goldstein}},\ and\ \bibinfo {author} {\bibfnamefont {R.}~\bibnamefont
  {Ilan}},\ }\bibfield  {title} {{\selectlanguage {english}\bibinfo {title}
  {Berry connection induced anomalous wave-packet dynamics in non-{Hermitian}
  systems}},\ }\href {https://doi.org/10.1103/PhysRevB.102.245147} {\bibfield
  {journal} {\bibinfo  {journal} {Physical Review B}\ }\textbf {\bibinfo
  {volume} {102}},\ \bibinfo {pages} {245147} (\bibinfo {year}
  {2020})}\BibitemShut {NoStop}%
\bibitem [{\citenamefont {Solnyshkov}\ \emph {et~al.}(2021)\citenamefont
  {Solnyshkov}, \citenamefont {Leblanc}, \citenamefont {Bessonart},
  \citenamefont {Nalitov}, \citenamefont {Ren}, \citenamefont {Liao},
  \citenamefont {Li},\ and\ \citenamefont
  {Malpuech}}]{solnyshkov_quantum_2021}%
  \BibitemOpen
  \bibfield  {author} {\bibinfo {author} {\bibfnamefont {D.~D.}\ \bibnamefont
  {Solnyshkov}}, \bibinfo {author} {\bibfnamefont {C.}~\bibnamefont {Leblanc}},
  \bibinfo {author} {\bibfnamefont {L.}~\bibnamefont {Bessonart}}, \bibinfo
  {author} {\bibfnamefont {A.}~\bibnamefont {Nalitov}}, \bibinfo {author}
  {\bibfnamefont {J.}~\bibnamefont {Ren}}, \bibinfo {author} {\bibfnamefont
  {Q.}~\bibnamefont {Liao}}, \bibinfo {author} {\bibfnamefont {F.}~\bibnamefont
  {Li}},\ and\ \bibinfo {author} {\bibfnamefont {G.}~\bibnamefont {Malpuech}},\
  }\bibfield  {title} {\bibinfo {title} {Quantum metric and wave packets at
  exceptional points in non-{Hermitian} systems},\ }\href
  {https://doi.org/10.1103/PhysRevB.103.125302} {\bibfield  {journal} {\bibinfo
   {journal} {Physical Review B}\ }\textbf {\bibinfo {volume} {103}},\ \bibinfo
  {pages} {125302} (\bibinfo {year} {2021})}\BibitemShut {NoStop}%
\bibitem [{\citenamefont {Schomerus}(2020)}]{schomerus_nonreciprocal_2020}%
  \BibitemOpen
  \bibfield  {author} {\bibinfo {author} {\bibfnamefont {H.}~\bibnamefont
  {Schomerus}},\ }\bibfield  {title} {\bibinfo {title} {Nonreciprocal response
  theory of non-{Hermitian} mechanical metamaterials: {Response} phase
  transition from the skin effect of zero modes},\ }\href
  {https://doi.org/10.1103/PhysRevResearch.2.013058} {\bibfield  {journal}
  {\bibinfo  {journal} {Physical Review Research}\ }\textbf {\bibinfo {volume}
  {2}},\ \bibinfo {pages} {013058} (\bibinfo {year} {2020})}\BibitemShut
  {NoStop}%
\bibitem [{\citenamefont {Miri}\ and\ \citenamefont
  {Alù}(2019)}]{AliMiriExceptional}%
  \BibitemOpen
  \bibfield  {author} {\bibinfo {author} {\bibfnamefont {M.-A.}\ \bibnamefont
  {Miri}}\ and\ \bibinfo {author} {\bibfnamefont {A.}~\bibnamefont {Alù}},\
  }\bibfield  {title} {\bibinfo {title} {Exceptional points in optics and
  photonics},\ }\href {https://doi.org/10.1126/science.aar7709} {\bibfield
  {journal} {\bibinfo  {journal} {Science}\ }\textbf {\bibinfo {volume}
  {363}},\ \bibinfo {pages} {eaar7709} (\bibinfo {year} {2019})},\ \Eprint
  {https://arxiv.org/abs/https://www.science.org/doi/pdf/10.1126/science.aar7709}
  {https://www.science.org/doi/pdf/10.1126/science.aar7709} \BibitemShut
  {NoStop}%
\bibitem [{\citenamefont {Dattoli}\ \emph {et~al.}(1990)\citenamefont
  {Dattoli}, \citenamefont {Mignani},\ and\ \citenamefont
  {Torre}}]{GDattoli_1990}%
  \BibitemOpen
  \bibfield  {author} {\bibinfo {author} {\bibfnamefont {G.}~\bibnamefont
  {Dattoli}}, \bibinfo {author} {\bibfnamefont {R.}~\bibnamefont {Mignani}},\
  and\ \bibinfo {author} {\bibfnamefont {A.}~\bibnamefont {Torre}},\ }\bibfield
   {title} {\bibinfo {title} {Geometrical phase in the cyclic evolution of
  non-hermitian systems},\ }\href {https://doi.org/10.1088/0305-4470/23/24/020}
  {\bibfield  {journal} {\bibinfo  {journal} {Journal of Physics A:
  Mathematical and General}\ }\textbf {\bibinfo {volume} {23}},\ \bibinfo
  {pages} {5795} (\bibinfo {year} {1990})}\BibitemShut {NoStop}%
\bibitem [{\citenamefont {Longhi}\ and\ \citenamefont
  {Feng}(2023)}]{longhi_complex_2023}%
  \BibitemOpen
  \bibfield  {author} {\bibinfo {author} {\bibfnamefont {S.}~\bibnamefont
  {Longhi}}\ and\ \bibinfo {author} {\bibfnamefont {L.}~\bibnamefont {Feng}},\
  }\bibfield  {title} {\bibinfo {title} {Complex {Berry} phase and imperfect
  non-{Hermitian} phase transitions},\ }\href
  {https://doi.org/10.1103/PhysRevB.107.085122} {\bibfield  {journal} {\bibinfo
   {journal} {Physical Review B}\ }\textbf {\bibinfo {volume} {107}},\ \bibinfo
  {pages} {085122} (\bibinfo {year} {2023})}\BibitemShut {NoStop}%
\bibitem [{\citenamefont {Zhang}\ \emph {et~al.}(2019)\citenamefont {Zhang},
  \citenamefont {Wang},\ and\ \citenamefont {Gong}}]{ZhangPhysRevA.99.042104}%
  \BibitemOpen
  \bibfield  {author} {\bibinfo {author} {\bibfnamefont {D.-J.}\ \bibnamefont
  {Zhang}}, \bibinfo {author} {\bibfnamefont {Q.-H.}\ \bibnamefont {Wang}},\
  and\ \bibinfo {author} {\bibfnamefont {J.}~\bibnamefont {Gong}},\ }\bibfield
  {title} {\bibinfo {title} {Quantum geometric tensor in
  $\mathcal{PT}$-symmetric quantum mechanics},\ }\href
  {https://doi.org/10.1103/PhysRevA.99.042104} {\bibfield  {journal} {\bibinfo
  {journal} {Phys. Rev. A}\ }\textbf {\bibinfo {volume} {99}},\ \bibinfo
  {pages} {042104} (\bibinfo {year} {2019})}\BibitemShut {NoStop}%
\bibitem [{\citenamefont {Zhu}\ \emph {et~al.}(2021)\citenamefont {Zhu},
  \citenamefont {Zheng}, \citenamefont {Zhu},\ and\ \citenamefont
  {Palumbo}}]{ZhuPhysRevB.104.205103}%
  \BibitemOpen
  \bibfield  {author} {\bibinfo {author} {\bibfnamefont {Y.-Q.}\ \bibnamefont
  {Zhu}}, \bibinfo {author} {\bibfnamefont {W.}~\bibnamefont {Zheng}}, \bibinfo
  {author} {\bibfnamefont {S.-L.}\ \bibnamefont {Zhu}},\ and\ \bibinfo {author}
  {\bibfnamefont {G.}~\bibnamefont {Palumbo}},\ }\bibfield  {title} {\bibinfo
  {title} {Band topology of pseudo-hermitian phases through tensor berry
  connections and quantum metric},\ }\href
  {https://doi.org/10.1103/PhysRevB.104.205103} {\bibfield  {journal} {\bibinfo
   {journal} {Phys. Rev. B}\ }\textbf {\bibinfo {volume} {104}},\ \bibinfo
  {pages} {205103} (\bibinfo {year} {2021})}\BibitemShut {NoStop}%
\bibitem [{\citenamefont {Sternheim}\ and\ \citenamefont
  {Walker}(1972)}]{sternheim_non-hermitian_1972}%
  \BibitemOpen
  \bibfield  {author} {\bibinfo {author} {\bibfnamefont {M.~M.}\ \bibnamefont
  {Sternheim}}\ and\ \bibinfo {author} {\bibfnamefont {J.~F.}\ \bibnamefont
  {Walker}},\ }\bibfield  {title} {\bibinfo {title} {Non-{Hermitian}
  {Hamiltonians}, {Decaying} {States}, and {Perturbation} {Theory}},\ }\href
  {https://doi.org/10.1103/PhysRevC.6.114} {\bibfield  {journal} {\bibinfo
  {journal} {Physical Review C}\ }\textbf {\bibinfo {volume} {6}},\ \bibinfo
  {pages} {114} (\bibinfo {year} {1972})}\BibitemShut {NoStop}%
\bibitem [{Note1()}]{Note1}%
  \BibitemOpen
  \bibinfo {note} {While non-Hermitian systems have already been shown to have
  corrections to the velocity even in 1-dimension for real electric fields
  \cite {silberstein_berry_2020}, we now understand them to arise due to a
  perturbative correction to the band energies, which is separate from the
  generalized anomalous velocity defined here.}\BibitemShut {Stop}%
\bibitem [{\citenamefont {Schomerus}(2022)}]{schomerus_fundamental_2022}%
  \BibitemOpen
  \bibfield  {author} {\bibinfo {author} {\bibfnamefont {H.}~\bibnamefont
  {Schomerus}},\ }\bibfield  {title} {\bibinfo {title} {Fundamental constraints
  on the observability of non-{Hermitian} effects in passive systems},\ }\href
  {https://doi.org/10.1103/PhysRevA.106.063509} {\bibfield  {journal} {\bibinfo
   {journal} {Physical Review A}\ }\textbf {\bibinfo {volume} {106}},\ \bibinfo
  {pages} {063509} (\bibinfo {year} {2022})}\BibitemShut {NoStop}%
\bibitem [{\citenamefont {Spring}\ \emph {et~al.}(2023)\citenamefont {Spring},
  \citenamefont {Könye}, \citenamefont {Gerritsma}, \citenamefont {Fulga},\
  and\ \citenamefont {Akhmerov}}]{spring_phase_2023}%
  \BibitemOpen
  \bibfield  {author} {\bibinfo {author} {\bibfnamefont {H.}~\bibnamefont
  {Spring}}, \bibinfo {author} {\bibfnamefont {V.}~\bibnamefont {Könye}},
  \bibinfo {author} {\bibfnamefont {F.~A.}\ \bibnamefont {Gerritsma}}, \bibinfo
  {author} {\bibfnamefont {I.~C.}\ \bibnamefont {Fulga}},\ and\ \bibinfo
  {author} {\bibfnamefont {A.~R.}\ \bibnamefont {Akhmerov}},\ }\href
  {http://arxiv.org/abs/2301.07370} {{\selectlanguage {english}\bibinfo {title}
  {Phase transitions of wave packet dynamics in disordered non-{Hermitian}
  systems}}} (\bibinfo {year} {2023})\BibitemShut {NoStop}%
\bibitem [{\citenamefont {Yao}\ and\ \citenamefont
  {Wang}(2018)}]{YaoShunyuPhysRevLett.121.086803}%
  \BibitemOpen
  \bibfield  {author} {\bibinfo {author} {\bibfnamefont {S.}~\bibnamefont
  {Yao}}\ and\ \bibinfo {author} {\bibfnamefont {Z.}~\bibnamefont {Wang}},\
  }\bibfield  {title} {\bibinfo {title} {Edge states and topological invariants
  of non-hermitian systems},\ }\href
  {https://doi.org/10.1103/PhysRevLett.121.086803} {\bibfield  {journal}
  {\bibinfo  {journal} {Phys. Rev. Lett.}\ }\textbf {\bibinfo {volume} {121}},\
  \bibinfo {pages} {086803} (\bibinfo {year} {2018})}\BibitemShut {NoStop}%
\bibitem [{\citenamefont {Yokomizo}\ and\ \citenamefont
  {Murakami}(2019)}]{YokomizoPhysRevLett.123.066404}%
  \BibitemOpen
  \bibfield  {author} {\bibinfo {author} {\bibfnamefont {K.}~\bibnamefont
  {Yokomizo}}\ and\ \bibinfo {author} {\bibfnamefont {S.}~\bibnamefont
  {Murakami}},\ }\bibfield  {title} {\bibinfo {title} {Non-bloch band theory of
  non-hermitian systems},\ }\href
  {https://doi.org/10.1103/PhysRevLett.123.066404} {\bibfield  {journal}
  {\bibinfo  {journal} {Phys. Rev. Lett.}\ }\textbf {\bibinfo {volume} {123}},\
  \bibinfo {pages} {066404} (\bibinfo {year} {2019})}\BibitemShut {NoStop}%
\bibitem [{\citenamefont {Rapoport}\ and\ \citenamefont
  {Goldstein}(2023)}]{RapoportPhysRevB.107.085117}%
  \BibitemOpen
  \bibfield  {author} {\bibinfo {author} {\bibfnamefont {O.}~\bibnamefont
  {Rapoport}}\ and\ \bibinfo {author} {\bibfnamefont {M.}~\bibnamefont
  {Goldstein}},\ }\bibfield  {title} {\bibinfo {title} {Generalized topological
  bulk-edge correspondence in bulk-hermitian continuous systems with
  non-hermitian boundary conditions},\ }\href
  {https://doi.org/10.1103/PhysRevB.107.085117} {\bibfield  {journal} {\bibinfo
   {journal} {Phys. Rev. B}\ }\textbf {\bibinfo {volume} {107}},\ \bibinfo
  {pages} {085117} (\bibinfo {year} {2023})}\BibitemShut {NoStop}%
\bibitem [{\citenamefont {Shindou}\ and\ \citenamefont
  {Imura}(2005)}]{SHINDOU2005399}%
  \BibitemOpen
  \bibfield  {author} {\bibinfo {author} {\bibfnamefont {R.}~\bibnamefont
  {Shindou}}\ and\ \bibinfo {author} {\bibfnamefont {K.-I.}\ \bibnamefont
  {Imura}},\ }\bibfield  {title} {\bibinfo {title} {Noncommutative geometry and
  non-abelian berry phase in the wave-packet dynamics of bloch electrons},\
  }\href {https://doi.org/https://doi.org/10.1016/j.nuclphysb.2005.05.019}
  {\bibfield  {journal} {\bibinfo  {journal} {Nuclear Physics B}\ }\textbf
  {\bibinfo {volume} {720}},\ \bibinfo {pages} {399} (\bibinfo {year}
  {2005})}\BibitemShut {NoStop}%
\bibitem [{\citenamefont {Culcer}\ \emph {et~al.}(2005)\citenamefont {Culcer},
  \citenamefont {Yao},\ and\ \citenamefont {Niu}}]{CulcerPhysRevB.72.085110}%
  \BibitemOpen
  \bibfield  {author} {\bibinfo {author} {\bibfnamefont {D.}~\bibnamefont
  {Culcer}}, \bibinfo {author} {\bibfnamefont {Y.}~\bibnamefont {Yao}},\ and\
  \bibinfo {author} {\bibfnamefont {Q.}~\bibnamefont {Niu}},\ }\bibfield
  {title} {\bibinfo {title} {Coherent wave-packet evolution in coupled bands},\
  }\href {https://doi.org/10.1103/PhysRevB.72.085110} {\bibfield  {journal}
  {\bibinfo  {journal} {Phys. Rev. B}\ }\textbf {\bibinfo {volume} {72}},\
  \bibinfo {pages} {085110} (\bibinfo {year} {2005})}\BibitemShut {NoStop}%
\bibitem [{\citenamefont {Balian}(1989)}]{balian_relation_1989}%
  \BibitemOpen
  \bibfield  {author} {\bibinfo {author} {\bibfnamefont {R.}~\bibnamefont
  {Balian}},\ }\bibfield  {title} {{\selectlanguage {english}\bibinfo {title}
  {Relation between position and quasi-momentum operators in band theory}},\
  }\href {https://doi.org/10.1051/jphys:0198900500180262900} {\bibfield
  {journal} {\bibinfo  {journal} {Journal de Physique}\ }\textbf {\bibinfo
  {volume} {50}},\ \bibinfo {pages} {2629} (\bibinfo {year}
  {1989})}\BibitemShut {NoStop}%
\bibitem [{\citenamefont {Jiang}\ \emph {et~al.}(2005)\citenamefont {Jiang},
  \citenamefont {Li}, \citenamefont {Zhang},\ and\ \citenamefont
  {Liu}}]{jiang_semiclassical_2005}%
  \BibitemOpen
  \bibfield  {author} {\bibinfo {author} {\bibfnamefont {Z.~F.}\ \bibnamefont
  {Jiang}}, \bibinfo {author} {\bibfnamefont {R.~D.}\ \bibnamefont {Li}},
  \bibinfo {author} {\bibfnamefont {S.-C.}\ \bibnamefont {Zhang}},\ and\
  \bibinfo {author} {\bibfnamefont {W.~M.}\ \bibnamefont {Liu}},\ }\bibfield
  {title} {{\selectlanguage {english}\bibinfo {title} {Semiclassical time
  evolution of the holes from {Luttinger} {Hamiltonian}}},\ }\href
  {https://doi.org/10.1103/PhysRevB.72.045201} {\bibfield  {journal} {\bibinfo
  {journal} {Physical Review B}\ }\textbf {\bibinfo {volume} {72}},\ \bibinfo
  {pages} {045201} (\bibinfo {year} {2005})}\BibitemShut {NoStop}%
\bibitem [{\citenamefont {Zawadzki}\ and\ \citenamefont
  {Rusin}(2011)}]{zawadzki_zitterbewegung_2011}%
  \BibitemOpen
  \bibfield  {author} {\bibinfo {author} {\bibfnamefont {W.}~\bibnamefont
  {Zawadzki}}\ and\ \bibinfo {author} {\bibfnamefont {T.~M.}\ \bibnamefont
  {Rusin}},\ }\bibfield  {title} {{\selectlanguage {english}\bibinfo {title}
  {Zitterbewegung (trembling motion) of electrons in semiconductors: a
  review}},\ }\href {https://doi.org/10.1088/0953-8984/23/14/143201} {\bibfield
   {journal} {\bibinfo  {journal} {Journal of Physics: Condensed Matter}\
  }\textbf {\bibinfo {volume} {23}},\ \bibinfo {pages} {143201} (\bibinfo
  {year} {2011})}\BibitemShut {NoStop}%
\bibitem [{\citenamefont {Lock}(1979)}]{Lock10.1119/1.11697}%
  \BibitemOpen
  \bibfield  {author} {\bibinfo {author} {\bibfnamefont {J.~A.}\ \bibnamefont
  {Lock}},\ }\bibfield  {title} {\bibinfo {title} {{The Zitterbewegung of a
  free localized Dirac particle}},\ }\href {https://doi.org/10.1119/1.11697}
  {\bibfield  {journal} {\bibinfo  {journal} {American Journal of Physics}\
  }\textbf {\bibinfo {volume} {47}},\ \bibinfo {pages} {797} (\bibinfo {year}
  {1979})}\BibitemShut {NoStop}%
\end{thebibliography}%

\end{document}